\documentclass[nofootinbib,twocolumn,preprintnumbers]{revtex4-2}
\pdfoutput=1
\usepackage{amsmath,amsthm,amssymb,multirow,psfrag}
\usepackage{epsfig}
\usepackage{color}
\usepackage{slashed}
\graphicspath{{./Figures/}}

\begin{document}

%%%%%%%%%%%new definitions: 
\def\lsim{\mathrel{\rlap{\lower4pt\hbox{\hskip1pt$\sim$}}
  \raise1pt\hbox{$<$}}}
\def\gsim{\mathrel{\rlap{\lower4pt\hbox{\hskip1pt$\sim$}}
  \raise1pt\hbox{$>$}}}
\newcommand{\vev}[1]{ \left\langle {#1} \right\rangle }
\newcommand{\bra}[1]{ \langle {#1} | }
\newcommand{\ket}[1]{ | {#1} \rangle }
\newcommand{\ev}{ {\rm eV} }
\newcommand{\kev}{{\rm keV}}
\newcommand{\mev}{{\rm MeV}}
\newcommand{\gev}{\mathrm{GeV}}
\newcommand{\tev}{{\rm TeV}}
\newcommand{\mpl}{$M_{Pl}$}
\newcommand{\mw}{$M_{W}$}
\newcommand{\Ft}{F_{T}}
\newcommand{\Zparity}{\mathbb{Z}_2}
\newcommand{\BLambda}{\boldsymbol{\lambda}}
\newcommand{\met}{\;\not\!\!\!{E}_T}
\newcommand{\beq}{\begin{equation}}
\newcommand{\eeq}{\end{equation}}
\newcommand{\bea}{\begin{eqnarray}}
\newcommand{\eea}{\end{eqnarray}}
\newcommand{\nn}{\nonumber}
\newcommand{\hc}{\mathrm{h.c.}}
\newcommand{\eps}{\epsilon}
\newcommand{\bwt}{\begin{widetext}}
\newcommand{\ewt}{\end{widetext}}
\newcommand{\draftnote}[1]{{\bf\color{blue} #1}}

\newcommand{\cO}{{\cal O}}
\newcommand{\cL}{{\cal L}}
\newcommand{\cM}{{\cal M}}

%References  
\newcommand{\fref}[1]{Fig.~\ref{fig:#1}} 
\newcommand{\eref}[1]{Eq.~\eqref{eq:#1}} 
\newcommand{\aref}[1]{Appendix~\ref{app:#1}}
\newcommand{\sref}[1]{Section~\ref{sec:#1}}
\newcommand{\tref}[1]{Table~\ref{tab:#1}}

\newcommand{\zg}{$h\rightarrow Z\gamma$}
\newcommand{\hgg}{$h\rightarrow \gamma\gamma$}
\newcommand{\hl}{$h\rightarrow 4\ell$}

%\title{\LARGE{{\bf The Intriguing Case Of Large $Z\gamma$: Exploring The Parameter Space Of Higgs Decays To 4 Leptons }}}
\title{\LARGE{On New Physics Contributions to the Higgs Decay to $Z\gamma$}}
\author{{\bf {Paul Archer-Smith$\,^{a}$, Daniel Stolarski$\,^{a}$, and Roberto Vega-Morales$\,^{b}$}}}

\affiliation{
$^a$Ottawa-Carleton  Institute  for  Physics,  Carleton  University,\\
1125  Colonel  By  Drive,  Ottawa,  ON  K1S  5B6,  Canada\\
$^b$ Departamento de F\'{i}sica Te\'{o}rica y del Cosmos and CAFPE,~Universidad de Granada, 
Campus de Fuentenueva, E-18071 Granada, Spain
}

\email{
PaulSmith3@cmail.carleton.ca \\
stolar@physics.carleton.ca\\
rvegamorales@ugr.es\\
}

\begin{abstract}
We explore models of new physics that can give rise to large (100\% or more) enhancements to the rate of Higgs decay to $Z\gamma$ while still being consistent with other measurements. We show that this is impossible in simple models with one additional multiplet and also in well motivated models such as the MSSM and folded SUSY. We do find models with several multiplets that carry electroweak charge where such an enhancement is possible, but they require destructive interference effects. We also show that kinematic measurements in Higgs decay to four leptons can be sensitive to such models. Finally we explore the sensitivity of four lepton measurements to supersymmetric models and find that while the measurement is difficult with the high luminosity LHC, it may be possible with a future high energy hadron collider.

%We explore Higgs decay processes to dibosons and four leptons ($h \rightarrow V_1 V_2$ and $h \rightarrow 4l$), with a specific focus on models with a large Higgs to Z-gamma contribution. We demonstrate that realistic models with an additional NP singlet and triplet can saturate current limits on the Higgs to Z-gamma process while still respecting existing limits on the Higgs to diphoton rate. The number of events required for discovery for this class of models, using four-lepton analysis, is also estimated. For comparison purposes, the same analysis is conducted on both the minimal supersymmetric standard model (MSSM) and folded supersymmetry (FSUSY). 
\end{abstract}

\maketitle

%%%%%%%%%%%%%%%%%%%%%%%%%%%%%
\section{Introduction} 
\label{sec:intro} 

%In a sense, all is right in the particle physics world. The Standard Model (SM) has held up in trial after trial, measurements again and again converge towards the predicted values, and physicists are left with the disconcerting sense that maybe the world is simply a non-natural reality governed by the sick whims of an ``anthropic principle".
%
%But we want answers and sitting pretty on the SM isn't going to get us anywhere closer to a theory of everything. With that out of the way, let's get going!

The Higgs boson's interactions with the electroweak gauge bosons, $W$, $Z$, $\gamma$ are now well established. Decays to $WW^*$~\cite{Aaboud:2018jqu,Aad:2019lpq, Sirunyan:2018egh, Sirunyan:2020tzo}, $ZZ^*$~\cite{Sirunyan:2018sgc, Aad:2013wqa} and $\gamma\gamma$~\cite{Aaboud:2018xdt} have all been measured and are consistent with Standard Model (SM) expectations at the $\mathcal{O}(20\%)$ level. Production of the Higgs in association with a $W$ or $Z$~\cite{Aad:2019lpq, Sirunyan:2018egh} is also consistent with the SM. One decay mode that, as of yet, has not been detected is $h\rightarrow Z\gamma$.
The most recent searches for this channel from ATLAS~\cite{Aad:2020plj} and CMS~\cite{Sirunyan:2018tbk} place limits on the branching ratio relative to the SM prediction. The strongest limit comes from ATLAS which sets an upper upper bound on the \zg~signal strength at 3.6 times the Standard Model prediction. 

%Experimental references \cite{Aad:2020plj,Aaboud:2017uhw,Aaboud:2016trl,Aad:2014fia,Sirunyan:2018tbk,Khachatryan:2014kca,Chatrchyan:2013vaa}

The SM contribution to this decay is known at leading order (one-loop)~\cite{Cahn:1978nz,Bergstrom:1985hp} as well as higher order QCD corrections~\cite{Spira:1991tj,Bonciani:2015eua,Gehrmann:2015dua}. One-loop contributions from generic NP models are known~\cite{Hue:2017cph} as well as contributions in SM effective field theory~\cite{Bellazzini:2018paj,Dedes:2019bew}. Phenomenological studies in specific models have also been carried out~~\cite{Weiler:1988xn,Chiang:2012qz,Cao:2013ur,Maru:2013bja,Yue:2013qba,Belanger:2014roa,Fontes:2014xva,Hammad:2015eca,Hung:2019jue,Liu:2020nsm,He:2020suf,Dev:2013ff,Cao:2018cms,Bizot:2015zaa}, as well as studies of this channel in the boosted regime~\cite{No:2016ezr}, and explorations of how lepton colliders can probe the effective coupling of $hZ\gamma$ in Higgs production~\cite{Cao:2015iua}.

In this work we explore the space of new physics (NP) models that can give large contributions to \zg, close to the current experimental upper bound, without being in conflict with all other Higgs measurements. Due to precise measurements of the properties of the Higgs, and especially the $Z$ and $\gamma$, we assume these fields are SM-like and that the NP contribution to \zg~arises at one loop due to new particles that couple to the Higgs and have electroweak quantum numbers. 
The SM contributions to \hgg{}~\cite{Ellis:1975ap,shifman1979low} and \zg{}~\cite{Cahn:1978nz,Bergstrom:1985hp} are dominated by one-loop contributions (in contrast to $h\rightarrow ZZ$, which is dominated by tree-level contributions), so one-loop NP can make $\mathcal{O}(1)$ modifications to the rate. 
Even in this case, however, the strong constraints on \hgg{} make it difficult to engineer such models, and we show below that it is impossible to achieve this goal in models with a single new scalar multiplet. Therefore, if large deviations to \zg{} are detected in the near future, this implies the existence of a complicated NP sector. 

\begin{figure*}[tb]
\centering
\begin{minipage}[c]{\textwidth}
\includegraphics[width=0.6\textwidth]{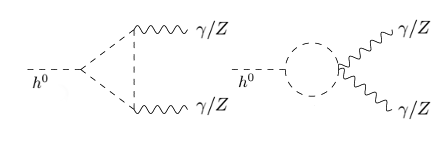}
\end{minipage}
\hfill
\caption{NP scalar one-loop contributions to Higgs to neutral diboson decays. The new scalar is represented by the unlabeled dashed lines. 
}
\label{fig:BSMContros}
\end{figure*}

We will focus on NP scalars $\Phi$ with hypercharge and/or $SU(2)_L$ charge that contribute to \zg. The case of other spins for the intermediate particles will not produce qualitatively different results. For a scalar $\Phi$, regardless of its quantum numbers, one can always write renormalizable quartic interactions:
\begin{equation}
\Phi^\dagger\Phi H^\dagger H,
\end{equation}
where $H$ is the SM Higgs doublet.\footnote{We use $h$ to denote the physical Higgs scalar and $H$ to denote the scalar doublet} These operators induce one-loop contributions to \hgg{},\,\,\zg{}, and $h\rightarrow ZZ$ of the type shown in Fig.~\ref{fig:BSMContros}. Trilinear operators written schematically as $H\Phi H$ are also possible for certain quantum numbers of $\Phi$, but these induce a vev for $\Phi$ which would give a significant contribution to the $\rho$ parameter~\cite{Ross:1975fq,Veltman:1977kh,Zyla:2020zbs} that is excluded. With several multiplets with different quantum numbers, more complicated structures are possible: these will be discussed in Sec.~\ref{sec:LargeZgamma}.

New scalars with electroweak quantum numbers can be produced at hadron colliders such as the LHC and searched for directly, but the limits will depend strongly on their decay modes. The analysis here, however, is more model independent and only depends on the quantum numbers, the mass, and the coupling of the Higgs to the new states. At lepton colliders such as LEP, because of the relatively low background rates and fully known beam parameters, searches can be done in a more model independent manner~\cite{Braibant:2003px,Heister:2002hp,Abdallah:2003xe,Achard:2003ge,Abbiendi:2002mp}; these exclude masses below about 105 GeV. Given this lower limit, the Higgs cannot decay to an on-shell $\Phi$, so we can expand the amplitudes from the diagrams in Fig.~\ref{fig:BSMContros} in powers of $m_h^2/4 m_\Phi^2$. The leading term generated will be a $CP$-even dimension five operator:
\begin{equation}
h F^{\mu\nu} F_{\mu\nu}
\label{eq:hff}
\end{equation}
where $F$ is the field strength tensor of the either the $Z$ or $\gamma$. 

\begin{figure*}[tb]
\centering
\begin{minipage}[c]{\textwidth}
\includegraphics[width=0.35\textwidth]{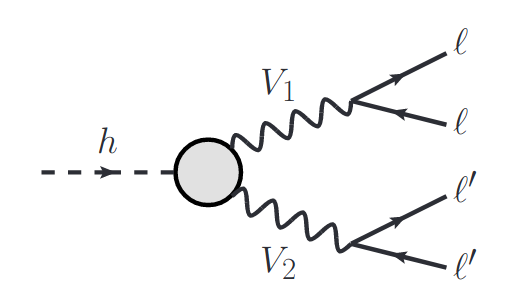}
\end{minipage}
\hfill
\caption{Representation of the $hVV$ corrections to the $h \rightarrow 4l$ amplitude where $V_{1,2} = Z, \gamma$ and $\ell, \ell' = e, \mu$.
}
\label{fig:hTo4l}
\end{figure*}

Operators of the type given in Eq.~\eqref{eq:hff} also contribute to the Higgs decay to four leptons, the so-called golden channel, via diagrams of the form shown in Fig.~\ref{fig:hTo4l}. The leading contribution to this decay is the tree-level contribution mediated by $ZZ^*$. Because this decay has four final state particles that can all be measured precisely, the rich final state kinematics can be used to gain significant information from each event. This means that one-loop contributions such as those from operators of the type in Eq.~\eqref{eq:hff} can be probed~\cite{Bolognesi:2012mm,Stolarski:2012ps,Chen:2013ejz,Chen:2014gka} within the lifetime of the LHC. This has been used for various applications including the probing of exotic light states~\cite{Falkowski:2014ffa}, measurement of the CP properties of the top Yukawa coupling~\cite{Chen:2015rha}, measurement of the ratio of the Higgs coupling to $WW$ relative to $ZZ$~\cite{Chen:2016ofc}, and the probing of various operators in SM effective field theory~\cite{Boselli:2017pef,Anderson:2013afp,Gainer:2014hha} as well as non-linear Higgs effects \cite{Liu:2019rce}. Even with the relatively small number of events already collected by the LHC, experiments can already use this data to place constraints on various scenarios~\cite{Sirunyan:2017tqd}.

Given models with large contributions to \zg{}, we can estimate how well those models can be probed in $h\rightarrow 4\ell$ as a function of the number of events using the analysis techniques in~\cite{Chen:2014gka,Chen:2015iha}. We use a simple hypothesis testing procedure on a few benchmark points of these models and find that the high luminosity run of the LHC can reach approximately $2\sigma$ sensitivity to these models in \hl.

Rather than adding arbitrary new scalar multiplets, one can also ask if well motivated models can produce significant deviations in \zg. Supersymmetric models are extremely well studied~\cite{Martin:1997ns}, and they contain scalar partners of the top quark, stops, which have large couplings to the Higgs and carry electroweak charges, so they can be scalars of the type shown in Fig.~\ref{fig:BSMContros}. The stops carry colour and, as such, contribute to production of the Higgs via gluon fusion in addition to the Higgs to diboson decays --- measurements of these processes (especially \hgg) can be used to place constraints on stop parameter space~\cite{Fan:2014txa}. Here we update the constraints from~\cite{Fan:2014txa} and show that these constraints imply that stops can only make small modifications to \zg. We also find that \hl{} at the LHC will not be sensitive to these models, but there may be sensitivity at future higher energy hadron colliders.

Another well motivated model is folded SUSY~\cite{Burdman:2006tz}, which also contains scalar top partners, F-stops, that have electroweak quantum numbers but \textit{do not} carry colour. This makes direct bounds much weaker, and it also eliminates constraints from Higgs production via gluon fusion, making indirect bounds also weaker~\cite{Fan:2014txa}. We also update the analysis on F-stops. While their contributions to \zg{} can be larger than for ordinary stops, it still cannot be large, and the conclusions for the four lepton analysis is similar to that of stops. 

% Additionally, we explore the parameter space of two supersymmetric (SUSY) models: the minimal supersymmetric standard model (MSSM) and folded SUSY \cite{Burdman:2006tz}. This is done for two primary reasons: 1) to demonstrate the relative ease of probing the large $Z\gamma$ space and 2) to show the limitations of four-lepton analysis within the SUSY parameter space.

The rest of the paper is organized as follows: Sec.~\ref{sec:Higgs} establishes the conventions used to describe new physics and the procedure used in analyzing the four-lepton processes, and Sec.~\ref{sec:LargeZgamma} presents the construction and signals of models with large contributions to \zg. Sec.~\ref{sec:SUSY} presents an analysis of supersymmetric models and Sec.~\ref{sec:Conclusion} wraps everything up.  

%Finally Appendix~\ref{app:SDTModel} explores a scenario with significant mass mixing between different electroweak multiplets.

%%%%%%%%%%%%%%%%%%%%%%%%%%%%%%%%
%%%%%%%%%%%%%%%%%%%%%%%%%%%%%%%%

\section{Higgs Physics}
\label{sec:Higgs}

Measurements of production and decay rates of the Higgs are all consistent with SM predictions. Therefore if there is new physics with electroweak charge that couples to the Higgs, it will be constrained by its contributions to the decay of Higgs to $\gamma\gamma$ at one loop. Since the leading SM contribution is also at one-loop, the constraints on such new physics at the weak scale will be strong. Similarly, new physics with colour charge that couples to the Higgs will contribute to the Higgs production via gluon fusion and can also place strong constraints. 

Assuming the Higgs width is small compared to its mass, the cross section of a specific production mode $i$ and decay to a specific final state $VV'$ can be parameterized as~\cite{Heinemeyer:2013tqa}: 
\begin{equation}\label{eqn:NWH}
\sigma(i \rightarrow h \rightarrow VV') = \frac{\sigma_i(\vec{\kappa})\Gamma^{VV'}(\vec{\kappa})}{\Gamma_H}.
\end{equation} 
Where $\Gamma_H$ is the total decay width of the Higgs, $\Gamma^{VV'}$ is the partial decay width to $VV'$, and $\sigma_i$ is the total cross section of $(i \rightarrow h \rightarrow \mathrm{Anything})$ in the $i$ production mode. Deviations from the SM can be parameterized through a set of coupling strength modifiers $\vec{\kappa}$~\cite{Aaboud:2018xdt}. For a given production process or decay mode $j$, these are defined by:
\begin{equation}\label{eqn:kappas}
\kappa^2_j = \frac{\sigma_j}{\sigma_{j,SM}} \,\,\,\,\, \mathrm{or} \,\,\,\,\, 
\kappa^2_j = \frac{\Gamma^j}{\Gamma^j_{SM}}.
\end{equation}  
The SM has all $\kappa_j$ values equal to unity. 
The $\kappa$ framework is not a full theory, and there are effects that it cannot account for that are detailed in~\cite{Heinemeyer:2013tqa}. The largest such effects are in processes where the Higgs is off-shell and not relevant for our analysis. We leave a more detailed accounting of the error budget due to this framework to future work. 

Generic NP contributions can be constrained by the limits placed by experimental measurements on Higgs boson coupling strength modifiers, $\kappa_j$. In the models we are considering where the new physics dominantly contributes at one loop, the relevant coupling modifiers are those to photons, gluons, and \zg: $\kappa_\gamma$ and $\kappa_g$, and $\kappa_{Z\gamma}$ respectively. The kinematics of the measured processes do not change significantly in the presence of new physics above half the Higgs mass~\citep{Aaboud:2018xdt}, so the $\kappa$ framework is sufficient to describe deviations from the SM. 

\begin{figure*}[tb]
\centering
\begin{minipage}[c]{\textwidth}
\includegraphics[width=.45\textwidth]{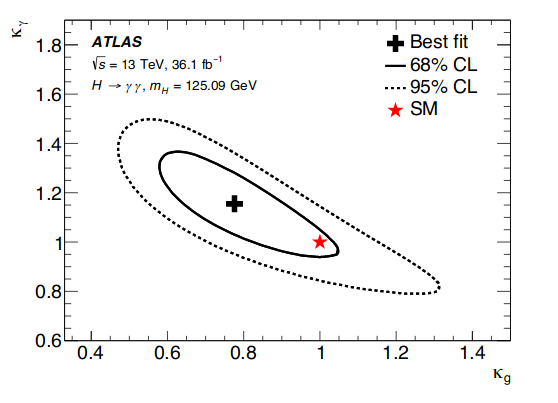}
\end{minipage}
\hfill
\caption{Likelihood contours in the $\kappa_{\gamma}$-$\kappa_{g}$ plane from ATLAS data~\cite{Aaboud:2018xdt}. All other coupling modifiers are fixed to the SM values.
}
\label{fig:aDPResults}
\end{figure*}

In Fig.~\ref{fig:aDPResults} we show the constraints in the $\kappa_\gamma-\kappa_g$ plane from~\cite{Aaboud:2018xdt} assuming all other couplings are SM-like. The best fit values resulting from their analysis are $\kappa_{\gamma} = 1.16^{+0.14}_{-0.14}$ and $\kappa_g = 0.76^{+0.17}_{-0.14}$. On the other hand, the upper bounds on the Higgs decay to $Z\gamma$ are, at 95\% confidence level, 3.6 times the Standard Model prediction for the production cross-section times the branching ratio for $pp \rightarrow h \rightarrow Z\gamma$  --- this drops to 2.6 times the SM prediction if the production is set to the SM value. The best fit value for the signal yield is $2.0^{+1.0}_{-0.9}$ \cite{Aad:2020plj}, normalized to the SM. The future HL-LHC is expected to improve these constraints for the $pp \rightarrow h \rightarrow Z\gamma$ process to $1.00 \pm 0.23$ the SM prediction~\cite{HLLHC}.
In this work we will explore models that are not excluded by the $\kappa_\gamma-\kappa_g$ analysis but that can have large contributions to \zg, possibly saturating the current experimental constraint of 2.6 times the SM prediction.  
%Ultimately, we use ATLAS' measurements to provide both constraints on large Z-gamma models in \ref{sec:LargeZgamma} and exclusion limits on supersymmetric parameter space in Sec. \ref{sec:SUSY}; in the latter case, the $Z\gamma$ limits are noncompetitive.

\subsection{\hl~at One Loop}

Despite the small rate, the \hl~mode benefits from a rich kinematic structure and very high signal to background ratio. This, coupled with systematic uncertainties that are very different (and typically smaller) than with direct diboson measurements, make the four-lepton channel an important complementary way to examine the Higgs sector. Here, we follow the anaylsis of~\cite{Chen:2014gka,Chen:2015iha} in order to constrain these models using their contribution to \hl~at one loop. The dominant contribution to \hl~ is via virtual gauge bosons with generic diagrams of the form of Fig.~\ref{fig:hTo4l}. 
%A central part of the $h \rightarrow 4l$ process is the connection of the NP to the vector bosons. This occurs either directly, in the case of the $hZZ$ tree level vertex, or indirectly through a loop, as is the case for the leading order terms for $h \rightarrow \gamma\gamma$ and $h \rightarrow Z\gamma$. 
The contributions to the $hVV$ couplings can be described by an effective Lagrangian:
\begin{equation}\label{eqn:effLagrangian}
\cL = \cL_0 + \cL_{1} + ...
\end{equation}
Here, the leading order term that contributes to the $h ZZ$ coupling at tree level is given in $\cL_0$. This term gifts the $Z$ particle its mass and is generated via EWSB,
\begin{equation}\label{eqn:TLZ}
\cL_0 = \frac{h}{2v}A_1^{ZZ}m_Z^2 Z^{\mu}Z_{\mu},
\end{equation}
where $Z_\mu$ is the $Z$ boson field. For the SM at tree level, we have $A_1^{ZZ} = 2$. Additionally, there are dimension-5 operators that can parameterize the one-loop contributions:
\begin{equation}\label{eqn:1LEff}
\cL_{1} = \frac{h}{4v}\left(A_2^{Z\gamma}F^{\mu\nu}Z_{\mu\nu} + A_2^{\gamma\gamma}F^{\mu\nu}F_{\mu\nu} +A_2^{ZZ}Z^{\mu\nu}Z_{\mu\nu}\right), \\
\end{equation}
%\begin{equation}\label{eqn:1LEff}
%\begin{split}
%\cL_{ZZ} = \frac{h}{4v}\left(A_2^{ZZ}Z^{\mu\nu}Z_{\mu\nu} + A_3^{ZZ}Z^{\mu\nu}\tilde{Z}_{\mu\nu}\right) \\
%\cL_{Z\gamma} = \frac{h}{2v}\left(A_2^{Z\gamma}F^{\mu\nu}Z_{\mu\nu} + A_3^{Z\gamma}F^{\mu\nu}\tilde{Z}_{\mu\nu}\right) \\
%\cL_{\gamma\gamma} = \frac{h}{4v}\left(A_2^{\gamma\gamma}F^{\mu\nu}F_{\mu\nu} + A_3^{\gamma\gamma}F^{\mu\nu}\tilde{F}_{\mu\nu}\right).
%\end{split}
%\end{equation}
where $Z_{\mu\nu}$ ($F_{\mu\nu}$) is the field strength of the $Z$ (photon).  These Lagrangian terms are only a subset of a more general description that would include other dimension five operators, but those operators are constrained to be too small to be relevant for this analysis~\cite{Gainer:2014hha,Chen:2015rha}. Moving forward, we neglect the contributions of $A_2^{ZZ}$ due to the lack of sensitivity \cite{Chen:2014pia,Chen:2014gka,Chen:2015iha} of future measurements to such modifications.

It should also be noted that in the SM there are one-loop box and pentagon diagrams that contribute to the \hl~process that are not captured by the parameterization of Eq.~\ref{eqn:1LEff}. 
%It should also be noted that the \hl~process also permits box and pentagon diagram contributions at one-loop in the cases where the Higgs directly couples to $ZZ$. 
These contributions are small, relative to the tree-level process, and are unaffected by the presence of any of the BSM models presented here. These calculations have been done \cite{Bredenstein:2006rh,Bredenstein:2006ha}, and we leave the integration of these results into our framework for future work. As such, we now turn our attention to the most relevant operators in the search for BSM: $A_2^{\gamma\gamma}$ and $A_2^{Z\gamma}$.
% such as $h\partial^{\mu}Z^{\nu}V_{\mu\nu}$, $hZ^{\mu}l\gamma_{\mu}l$, or $\square h Z^{\mu}Z_{\mu}$. The latter two operators are only relevant in the case of an off-shell NP \cite{Gainer:2014hha} and can thus be ignored for the purposes of this work. The first is generated at one-loop for $h\rightarrow Z\gamma$, but vanishes for an on-shell photon and also vanishes to leading order in the heavy loop particle expansion~\cite{Chen:2015rha}.

Within the SM the $A_2^{V\gamma}$ couplings will be generated primarily via a W boson loop along with (smaller) contributions from a top loop. 
%After taking these loops into account, the next most significant contribution comes from a bottom quark loop. This is heavily suppressed due to the comparative weakness of the bottom Yukawa coupling; the overall suppression is of order $\sim (m_b/m_t)^2$. With all other contributions being even smaller, we can safely treat the SM contributions to these effective couplings as coming exclusively through the W boson and top loops. 
Numerically, the SM values are $A_2^{\gamma\gamma} \approx - 0.008$ and $A_2^{Z\gamma} \approx -0.014$ \cite{Chen:2016ofc}.\footnote{Note that \cite{Low:2012rj} gives a different sign for $A_2^{Z\gamma}$.} 
%The $A_3^{V\gamma}$ coefficients, in contrast, are zero at this loop order and can be neglected for this analysis. 
%
These form factors can be used to compute on-shell Higgs decay to $\gamma\gamma$ or $Z\gamma$. These SM one-loop contributions have been explored in the literature for both \zg~\cite{Cahn:1978nz,Bergstrom:1985hp} and \hgg~\cite{Ellis:1975ap,shifman1979low}. 
We have assumed NP contributions are $CP$-even 
%All NP contributions examined within this paper only modify the $A_2$ couplings as we have assumed NP is parity conserving 
because of strong constraints from flavour and $CP$ violating observables~\cite{Brod:2013cka}.\footnote{For analyses of $CP$ violation in \zg~see~\cite{Farina:2015dua,Chen:2017plj,Chen:2014ona}.} 
%\begin{figure*}[tb]
%\centering
%\begin{minipage}[c]{\textwidth}
%\includegraphics[width=0.9\textwidth]{smContributions.png}
%\end{minipage}
%\hfill
%\caption{Standard Model one-loop contributions from top quarks (left) and W bosons for $h \rightarrow V_1 V_2 \rightarrow 4l$ \cite{Chen:2015rha}.
%}
%\label{fig:SMContros}
%\end{figure*}

%\subsection{Introducing BSM}

%\subsection{One-loop Contributions}

Given this parameterization, we express the matrix element for the $h \rightarrow 4l$ process in the form,
\begin{equation}\label{eqn:MatStruchTo4l}
\begin{split}
\cM(h\rightarrow 4l) = \cM^{\mu\nu}(h\rightarrow V_1 V_2)\times \,\,\,\,\,\,\,\,\,\,\,\,\,\,\,\,\,\,\,\,\,\,\,\,\,\,\,\,\,\,
\\
\,\,\,\,
\mathcal{P}_{\mu\alpha}(V_1)\cM^{\alpha}(V_1 \rightarrow 2l)\mathcal{P}_{\nu\beta}(V_2)\cM^{\beta}(V_2\rightarrow 2l),
\end{split}
\end{equation}
where $V = Z, \gamma$, and $\mathcal{P}_{\mu\nu}(V_i)$ are the propagators of the vector bosons. It should be noted that the first line is where any potential NP that we could hope to uncover would reside. The second line is simply vector bosons propagating and decaying to leptons --- processes that are well understood and well measured. The $h \rightarrow V_1V_2$ matrix element can be parameterized as follows:
\begin{equation}\label{eqn:tensorStruc}
\begin{split}
\cM^{\mu\nu}(h\rightarrow V_1 V_2) = \frac{1}{v}C_1^i m_z^2 g^{\mu\nu} + \,\,\,\,\,\,\,\,\,\,\,\,\,\,\,\,\,\,\,\,\,\,\,\,\,\,\,\,\,\,
\\
\frac{1}{v}C^i_2(k_1^{\nu}k_2^{\mu} - k_1 \cdot k_2 g^{\mu\nu}) + \frac{1}{v}C^i_3 \epsilon^{\mu\nu\alpha\beta}k_{1\alpha}k_{2\beta},
\end{split}
\end{equation}
with $i = ZZ, Z\gamma, \gamma\gamma$ and $k_1,\, k_2$ representing the four momenta of the intermediate vector bosons (or, equivalently, lepton pairs).

The form factors, $C^i_n$, are Lorentz invariant and encode the momentum dependence. They have the form, 
\begin{equation}\label{eqn:ffStruc}
C^i_n \sim h_X f_i(m_h^2/m_X^2, k_1^2, k_2^2),
\end{equation}
with $f_i(m_h^2/m_X^2, k_1^2, k_2^2)$ representing the loop function for NP particle $X$ coupled to the Higgs with coupling $h_X$. Previous studies have indicated that dependence on the invariant mass parameters $k_i^2$ is quite weak \cite{Chen:2015rha,Gonzalez-Alonso:2014eva}. In order to examine this claim, the form factors $C^i_n$ were expressed as a Taylor Series and expanded around the pole masses of the intermediate vector bosons. The parameter space examined here can indeed be probed taking $k_i^2 = m^2$ for the relevant boson, i.e.~treating the vector as on-shell for the form factor. This limit is used for determining the exclusion bounds from ATLAS measurements of Higgs to diphotons. 

All the NP scenarios we will explore can then be encoded in the value of the effective field theory (EFT) operators $A^{\gamma\gamma}_{2}$ and $A^{Z\gamma}_{2}$ for all models (plus the gluon form factor $A^{gg}_{2}$ for coloured new scalars such as stops). They are computed via the diagrams in Fig.~\ref{fig:BSMContros} assuming the intermediate state gauge bosons are on-shell --- in the language of Eq.~\ref{eqn:tensorStruc} this makes $A^{\gamma\gamma}_{2} = 2C^{\gamma\gamma}_2$ and $A^{\gamma\gamma}_{2} = 2C^{Z\gamma}_2$. 

\subsection{Kinematic Analysis of \hl}\label{sec:4l}

%The so-called ``golden channel" of NP physics is the process where after the NP decays to some combination of Zs and gammas, the two produced bosons decay into electron and muon pairs: $h \rightarrow V V \rightarrow 4l$. 
In the Higgs rest frame, the useful kinematic variables for this decay are:
\begin{itemize}
\item $\Phi$: The decay angle between the decay planes of the intermediate bosons in the rest frame of the Higgs.
\item $\theta_1$: The angle between the lepton coming from the decay of $V_1$ and the momentum of $V_2$ in the $V_1$ rest frame.
\item $\theta_2$: Identical to $\theta_1$ but with $V_1 $ and $V_2$ swapped.
\item $M_i$: The invariant mass of the lepton pair produced by the decay of the $i^{th}$ boson. By convention $M_1 > M_2$.  
\end{itemize}

In the case of intermediate $Z\gamma$ vectors, the rate is dominated by $M_1 \approx M_Z$. Kinematics require $M_1 + M_2 \leq \sqrt{s}$ where $s$ is the invariant mass squared of the four lepton system. We refer to the set of these variables for a single $4\ell$ event as $Y$ --- this will be used as the input for the likelihood analysis to test between the Standard Model and different NP scenarios, or equivalently as ways to test different values of the form factors $A^{VV'}_2$. 
%put into use through our likelihood analysis in Sec.\,\ref{sec:LargeZgamma} and Sec.\,\ref{sec:SUSY}.
For different values values of the form factors, we generate Monte Carlo (MC) events for the process $gg \rightarrow h \rightarrow 4l$ at the 14 TeV LHC using MadGraph5\texttt{\textunderscore}aMC$@$NLO~\cite{Alwall:2014hca}. Since gluon fusion is the primary production mechanism for the Higgs at the LHC~\cite{Dittmaier:2011ti} and the variables examined during our analysis are in the rest frame of the Higgs and thus only sensitive to the decay process, including other production modes would not change our analysis. Additionally, the very high signal to background ratio of the four lepton channel and the easy identification of leptons within LHC detectors makes both a full simulation of backgrounds and an examination of detector effects subleading for the purposes of this work \cite{Chen:2014pia}. Measurements (see for example~\cite{Sirunyan:2021rug}) in this channel confirm that the signal to background ratio is very large in the Higgs invariant mass window.
 
We now determine the number of events needed to distinguish two different hypotheses (SM vs.~NP) at a given confidence level. 
This is done with a likelihood analysis of the MC generated NP events compared  to MC events assuming purely SM physics. A standard unbinned likelihood analysis is used (see~\cite{Gao:2010qx, Stolarski:2012ps} for further details) where the computed differential cross-section can be normalized and used as a probability over the $4l$ kinematic variables: 
\begin{equation}
P(Y|A) = \frac{\left|\mathcal{M}(Y,A) \right|^2}{\int_{\rm fid} \left|\mathcal{M}(Y,A) \right|^2 \mathrm{dY}} \, ,
\end{equation}
where $Y$ is the set of kinematic variables of an event, $A$ is the given set of effective coupling strengths for the model being examined, and $\cM$ is the matrix element from Eq.\,\ref{eqn:MatStruchTo4l}. The integral is over the fiducial acceptance region for $4\ell$ events. If we have $N$ events, this allows us to compute a likelihood: $\mathcal{L} (A) = \prod^N_{i=1}P(Y_i|A)$. If we have two different scenarios (typically a SM scenario and a NP scenario), we can calculate a likelihood ratio and construct a hypothesis test statistic, 
\begin{equation}\label{eqn:testStat}
\Lambda = 2 \log \left(\mathcal{L}(A_1)/\mathcal{L}(A_2)\right).
\end{equation} 
This test statistic is calculated on MC data generated assuming each of the two underlying hypotheses. This creates two distributions of the test statistic $\Lambda$, one assuming scenario 1 is true, the other, scenario 2. The separation between these distributions is a measure of how easily distinguishable the two hypotheses are. Explicitly, if we call the distribution with the smaller average test statistic $f$ and the other distribution $g$ then there exists a value $\Lambda_0$ such that, 
\begin{equation}\label{eqn:testStatComp}
\int^{\Lambda_0}_{-\infty} f(\Lambda) \,\mathrm{d}\Lambda = 
\int_{\Lambda_0}^{\infty} g(\Lambda)\, \mathrm{d}\Lambda.
\end{equation} 
Because $\Lambda_0$ is the point at which the distributions are indistinguishable, the value on either side of Eq.\,\ref{eqn:testStatComp} is the one-sided Gaussian probability of the observed events of one scenario excluding the alternative. This can be converted into a more standard $\sigma$ value that is a function of the number of signal events $N$. In other words, we can get the expected statistical significance as a function of the number of events for any two scenarios.

%%%%%%%%%%%%%%%%%%%%%%%%%%%%%%%%
%%%%%%%%%%%%%%%%%%%%%%%%%%%%%%%%

\section{Models with large $h \rightarrow Z\gamma$ contributions}
\label{sec:LargeZgamma}

Here, we explore models with new electroweak multiplets that can give large contributions to \zg~while not being in conflict with \hgg.

\subsection{New Scalar Multiplets}
\label{sec:newScalar}

%When exploring possible NP scenarios that could lead to a large NP decay to $Z\gamma$, w
We begin with a simple scenario: a single scalar multiplet that is charged under $\mathrm{SU}(2) \times \mathrm{U}(1)$, but is a singlet under $\mathrm{SU}(3)_c$. At the moment, we remain agnostic to the size of the multiplet; we do note, however, that the size of the multiplet is bounded at 8 (isospin of 7/2) due to perturbative unitary constraints~\cite{Hally:2012pu}. Finally, in order to stay away from any LEP II bounds~\cite{Braibant:2003px,Heister:2002hp,Abdallah:2003xe,Achard:2003ge,Abbiendi:2002mp}, we require our particles have a mass $\gtrsim 100\, \mathrm{GeV}$.  

The NP contributions to our effective operators $A^{Z\gamma}_2$ and $A^{\gamma\gamma}_2$ have the form
\begin{equation}\label{eqn:LZGOppForm}
\begin{split}
A^{\gamma\gamma}_{2} = \sum_{\rm particles} 2 h_{X}g_{X}^2*l_{\gamma\gamma}[m]
\\
A^{Z\gamma}_2 =\sum_{\rm particles} 2 h_{X}g_{X}z_{X}*l_{Z\gamma}[m]
\end{split}
\end{equation}
where $h_X$ is the coupling of the new particles to the Higgs, $g_X$ and $z_X$ are the coupling of the new particles to photons and $Z$ bosons, respectively, and $l_{\gamma\gamma}$ and $l_{Z\gamma}$ are the loop functions that depend on the mass of the NP particles given in App.~\ref{app:loops}. The sum is over all states in the multiplet. It should be noted that, unless there is mass mixing, $h_X$ and the particle mass is the same for all members of the multiplet. The couplings $g_X$ and $z_X$, on the other hand, depend on the $T_3$ value for the particle and are thus different for every multiplet member. Additionally, for NP particles with mass $\gtrsim$ 100 GeV, the loop functions become approximately equal, $l_{\gamma\gamma}[m]\approx l_{Z\gamma}[m]$, so the relative contribution sizes of the new particle multiplet to the diphoton vs.~$Z\gamma$ is controlled by the relative size of 
\begin{eqnarray}\label{eqn:geff}
 g^{\rm eff}_{\gamma\gamma} &=& \sum g_X^2\nonumber \\
  g^{\rm eff}_{Z\gamma} &=& \sum g_X z_X,
\end{eqnarray}   
  with the sum being over all particles in the multiplet.  

The relevant terms in the Lagrangian for a single scalar multiplet $X$ are given by
\begin{equation}\label{eqn:SingletLagrangian}
%\begin{split}
\cL \supset (D_\mu X)^\dagger(D^{\mu} X) + m_X^2 X^\dagger X+ b (X^\dagger X)(H^\dagger H) \, ,
%\\
%+\, a (X^\dagger X)^2 
%\end{split}
\end{equation}
with $D_\mu$ being the typical covariant derivative that gives rise to the coupling of $X$ to gauge bosons. 
The coupling $b$ gives rise to the Higgs coupling to $X$ pairs after electroweak symmetry breaking, and we ignore $X$ self-interactions as they do not affect the analysis. We assume $m_X^2 > 0$, as bounds on new electroweak scalars with vacuum expectation values are very strong~\cite{Ross:1975fq,Veltman:1977kh,Zyla:2020zbs} (unless $X$ has the same quantum numbers as the Higgs). For the same reason, we do not include $HXH$ terms which are allowed for certain quantum numbers of $X$. Finally, we remain agnostic as to which decay modes are present for the new particles; this allows us to ignore direct search limits and focus wholly on Higgs decays.

After a rotation into the mass basis for the gauge bosons, the aforementioned couplings for the new scalars are:
\begin{equation}\label{eqn:ScalarBosonCouplings}
\begin{split}
g_X = g\,T_3\,s_{W} + g'\,Y\,c_{W}
\\
z_X = g\,T_3\,c_{W} - g'\,Y\,s_{W},
\end{split}
\end{equation}
where $g$ and $g'$ are the gauge couplings for $SU(2)_L\times U(1)_Y$ and $s_W$ and $c_W$ are the sine and cosine of the weak mixing angle. 
%The two parameters that we can control within this equation are the 3rd component of weak-isospin, $T_3$, and the weak hypercharge, $Y$. 
The typical convention relating electric charge of the scalars to the isospin and hypercharge holds: $Q = T_3 + Y$. Requiring integer charge and half-integer (integer) values for weak-isospin restrict hypercharge to half-integer (integer) values.

\subsection{The Failure Of One Additional Multiplet}

%With the foundation for our NP multiplet in place, 
We now demonstrate that the inclusion of a single multiplet cannot give rise to large contributions to \zg.
%the lack of large $Z\gamma$ solutions available within this simple framework. 
%We start by looking at individual terms within the sums of $g^{eff}_{\gamma\gamma}$ and $g^{eff}_{Z\gamma}$. Since the sum is over the weak-isospin value $T_3$, the effective couplings will be functions of the hypercharge for given multiplet sizes, 
The relevant products of couplings to gauge bosons are given by 
\begin{equation}\label{eqn:effCouplingPart}
\begin{split}
g_X^2 &= g^2\,T_3^2 s_W^2 + gg'T_3 c_W s_W\,Y + g'^2 c_W^2\,Y^2
\\
g_X z_X &= g^2\,T_3^2 s_W c_W + gg'T_3\left(c_W^2-s_W^2\right)\,Y - g'^2 c_W s_W\,Y^2.
\end{split}
\end{equation} 
%
%with $c_W$ and $s_W$ representing the cosine and sine of the Weinberg angle, respectively. 
Summing over $T_3$ gives effective couplings (defined in Eq.~\eqref{eqn:geff})
\begin{equation}\label{eqn:effCoupling}
\begin{split}
g^{eff}_{\gamma\gamma} &= K*(g^2\,\sin^2{\theta_W}) + k*(g'^2\cos^2{\theta_W}\,Y^2)
\\
g^{eff}_{Z\gamma} &= K*(g^2\sin{\theta_W}\cos{\theta_W}) - k*(g'^2\cos{\theta_W}\sin{\theta_W}\,Y^2)
\end{split}
\end{equation}  
where $k$ is the size of the multiplet and 
\begin{equation}\label{eqn:T3Series}
\begin{split}
K = 2\sum^{i = \frac{k-1}{2}}_i i^2 \,\,\, k \,\, \mathrm{odd}
\\
K = \frac{1}{2}\sum^{i = \frac{k}{2}}_i i^2 \,\,\, k \,\, \mathrm{even}.
\end{split}
\end{equation}
Since $g'^2\cos^2{\theta_W} > g'^2\cos{\theta_W}\sin{\theta_W}$, $g^{eff}_{\gamma\gamma}$ will always increase in strength faster than $g^{eff}_{Z\gamma}$ as the absolute value of the multiplet's hypercharge becomes larger. So, to maximize the relative size of the $Z\gamma$ contributions, the magnitude of the multiplet's hypercharge must be kept as small as possible
% --- this leads to a hypercharge of 0 for multiplets with odd numbers of particles and $Y=\pm1/2$ for those with an even number. In the case of odd multiplets, this gives us a best case scenario of 
Therefore, the best case scenario is an odd multiplet with $Y=0$ that gives
\begin{equation}\label{eqn:effCouplingRatio}
\frac{g^{eff}_{Z\gamma}}{g^{eff}_{\gamma\gamma}} = \cot{\theta_W} \approx 1.86. 
\end{equation} 
The even multiplets give weaker results, though as $k$ becomes larger, this ratio approaches the same value as the odd multiplets. When we consider the fact that in the SM the $Z\gamma$ effective operator is larger than the $\gamma\gamma$ one, 
\begin{equation}\label{eqn:A2Ratio}
\frac{A_2^{Z\gamma}}{A_2^{\gamma\gamma}}\Biggr\rvert_{\rm SM} \approx 1.75,
\end{equation}
 this shows the modification of to the \zg~and \hgg~rates are at best comparable. As such, the large \zg~parameter space cannot be probed by such simple models and we must add another degree of complexity through the addition of more multiplets. These scenarios can result in large \zg~rates through the cancellation of \hgg~contributions and/or with mass mixing effects such that the mass eigenstates with comparatively small $\gamma$ couplings have low masses and those with larger electric charges have larger masses. The mixing scenario turns out to be less compelling: even if one were to obtain mass mixing such that the member of the multiplet with the largest $T_3$ value (e.g. +1/2 for a doublet) that has the largest contribution to \hgg~relative to \zg~is effectively decoupled, the improvements to Eq.~\eqref{eqn:effCouplingRatio} are minor. Even with more complicated scenarios, it is extremely difficult to construct a model with a viable scalar potential that achieves the required mass splitting.
 With this in mind, for the rest of this section we explore cancellation scenarios.

\subsection{The Singlet-Triplet Model}\label{sec:singletTriplet}

The road map of where to go next comes from analyzing this question from the point of view of the unbroken electroweak effective field theory. In the EFT prior to spontaneous symmetry breaking~\cite{Weinberg:1979sa,Buchmuller:1985jz,Leung:1984ni} (typically called SMEFT) there are three dimension-6 terms relevant to Higgs to electroweak diboson decays~\cite{Grzadkowski:2010es}:
\begin{equation}\label{eqn:EFTOpps}
\begin{split}
\bullet &\,\, Q_{W} H^\dagger H W^{\mu\nu a}W^a_{\mu\nu}
\\
\bullet &\,\, Q_{B} H^\dagger H B^{\mu\nu}B_{\mu\nu}
\\
\bullet &\,\, Q_{WB} H^\dagger \tau^a H B^{\mu\nu}W^a_{\mu\nu}
\end{split}
\end{equation}
 where $W^{\mu\nu a}$ and $B^{\mu\nu}$ are the field strength tensors of $SU(2)_L$ and $U(1)_Y$ respectively, and $\tau^a$ are Pauli matrices. The introduction of scalar multiplets can generate contributions to the Wilson coefficients $Q$ of either sign: this can lead to multiplets that either enhance, dampen, or even nullify each other's effects. 
 
% These operators can be rotated into the mass basis via the relationships,
%
%\begin{equation}\label{eqn:GaugeBasisRot}
%\begin{split}
%B_\mu = \cos{\theta_W} A_\mu - \sin{\theta_W} Z_\mu
%\\
%W^3_\mu = \sin{\theta_W} A_\mu + \cos{\theta_W} Z_\mu
%\end{split}.
%\end{equation}

We can get expressions for our dimension-5 EFT in the Higgs basis defined in Eq.~\eqref{eqn:1LEff} in terms of our unbroken operators:
\begin{equation}\label{eqn:SMEFT}
\begin{split}
%A_2^{ZZ} = 8v^2\left(c_W^2 Q_{HW} + s_W^2 Q_{HB} +2s_W c_W Q_{HWB}\right) 
%\\
A_2^{Z\gamma} = 4v^2\left(s_W c_W (Q_{W} - Q_{B}) + (s_W^2 - c_W^2) Q_{WB}\right) 
\\
A_2^{\gamma\gamma} = 8v^2\left(s_W^2 Q_{W} + c_W^2 Q_{B} -2s_W c_W Q_{WB}\right).
\end{split}
\end{equation}
In the 3-dimensional space of the of $Q_W$, $Q_B$, and $Q_{WB}$, there is a plane of values that lead to no contribution to $A_2^{\gamma\gamma}$ --- yet the vast majority of this plane (everything outside one line) does offer up contributions to $A_2^{Z\gamma}$. 
%In essence, the same principle of nullification and enhancement mentioned above in the unbroken context can be applied to each multiplet's contribution to the effective operators $A_2^{\gamma\gamma}$ and $A_2^{Z\gamma}$ (Eq.\,\ref{eqn:LZGOppForm}). 
The task, then, becomes finding an appropriate set of multiplets that can simultaneously enhance the rate of $h \rightarrow Z\gamma$ and leave $h \rightarrow \gamma\gamma$ close to its SM value. 

A model that features these properties is one where we introduce two NP scalar multiplets: a singlet state, $S$, with $Y= 1$ and a triplet state, $T$, with $Y= 0$. Both multiplets are singlets under $\mathrm{SU}(3)_c$. At leading order, integrating out $S$ generates $Q_B$ and integrating out $T$ gives $Q_W$. Therefore, by choosing couplings appropriately, we can be on the line in the $Q_B-Q_W$ plane where the contributions to \hgg~(approximately) vanish. As we will consider relatively light states, we will use the EFT as only a guide and do our analysis in the full theory.

The relevant NP Lagrangian terms are:
\begin{equation}\label{eqn:STLagrangian}
\begin{split}
\cL \supset (D_\mu S)^\dagger(D^{\mu} S) + (D_\mu T)^\dagger(D^{\mu} T) + m_{S}^2 S^\dagger S 
\\
+\, m_{T}^2 T^\dagger T + \, a (S^\dagger S)^2 + b (S^\dagger S)(H^\dagger H) 
\\
 +\, c(T^\dagger T)^2  + d (T^\dagger T)(H^\dagger H) + e  (S^\dagger S)(T^\dagger T). 
\end{split}
\end{equation}
%
%Thankfully, most of these terms aren't relevant for the processes we're looking at. 
The contributions to the Higgs to diboson decays are simply those from Eq.\,\eqref{eqn:LZGOppForm} generalized to allow for multiple particles, 
\begin{equation}\label{eqn:STOppForm}
\begin{split}
A^{\gamma\gamma}_{2} = 2 (h_{S}g_{S}^2*l_{\gamma\gamma}[m_S] +h_{T}g^{eff}_{T\gamma\gamma}*l_{\gamma\gamma}[m_T] )
\\
A^{Z\gamma}_2 = 2(h_{S}g_{S}z_{S}*l_{Z\gamma}[m_S] + h_{T}g^{eff}_{TZ\gamma}*l_{Z\gamma}[m_T]), 
\end{split}
\end{equation}
where $h_S$ and $h_T$ are equal to $b\,v$ and $d\,v$, respectively; while $g^{eff}_{T\gamma\gamma}$ and $g^{eff}_{TZ\gamma}$ are calculated through Eq.\,\eqref{eqn:effCoupling}. Just as in the single NP multiplet case, the particle masses we consider are greater than $100$ GeV and, as a result, the loop functions for $\gamma\gamma$ and $Z\gamma$ are nearly identical. 
%Since $g_{S}^2$, $g^{eff}_{T\gamma\gamma}$, $g_{S}z_{S}$, and $g^{eff}_{TZ\gamma}$ are determined by the SU$(2) \times$U$(1)$ charges, Eq.\,\ref{eqn:STOppForm} reduces to a system of equations that can be solved for $h_S\,l[m_S]$ and $h_T\,l[m_T]$ for given NP contributions to the effective operators. 

Cancellation of the NP contributions to $A^{\gamma\gamma}_{2}$ requires either $h_S$ or $h_T$ (or, in terms of the Lagrangian, $b$ or $d$) to be negative. For concreteness we select $h_S$ to be negative. The scalar potential must be bounded from below in order to maintain its stability: this can be checked by ensuring that the quartic terms of the scalar potential have a limit $V_s \geq 0$ in all directions (any direction where $V_s = 0$ also requires the limit of the quadratic terms in that direction to be $\geq 0$). This requires the quartic couplings $a$ and $c$ to be positive, and we have also taken $d>0$. As it does not affect our analysis, we can also take $e>0$.  

The only remaining constraint is $b<0$ potentially leading to an unbounded direction. This analysis is nearly identical to those in 2 Higgs Doublet Models \cite{Branco:2011iw}. 
%Looking at directions where $H^\dagger H$ and $T^\dagger T$ are $0$ we see that $a$ is required to be positive. 
Setting $H^\dagger H = r \cos{\theta}$ and $S^\dagger S = r \sin{\theta}$, followed by taking the large $r$ limit gives
\begin{equation}\label{eqn:LargeRLimSP}
r^2 \,\left(a \sin^2{\theta} + \lambda \cos^2{\theta} + b \cos{\theta}\sin{\theta}\right) = r^2 f(\theta)
\end{equation}
where $\lambda$ is the SM quartic Higgs coupling. Ensuring Eq.\,\eqref{eqn:LargeRLimSP} is positive can be done by requiring $f(\theta) \geq 0$ at its smallest point. Applying this requirement enforces a minimum size on the negative coupling $b$: 
\begin{equation}\label{eqn:bRestriction}
b \geq -\sqrt{4a\lambda}.
\end{equation} 
The singlet-triplet model is representative of a much larger class of possible models --- not a unique solution to a large $Z\gamma$ signal.  This construction can apply to two multiplets of any sizes, although this particular choice of charges is simpler than the generic one because there are no allowed mass mixing terms.

Within this model, we will work with the following benchmark parameter point:
\begin{equation}\label{eqn:STParams}
\begin{split}
\bullet &\,\, m_S = 105\,\, \mathrm{GeV}
\\
\bullet &\,\, m_T = 140\,\, \mathrm{GeV}
\\
\bullet &\,\, a = 1.6 
\\
\bullet &\,\, b = -0.9 .
\end{split}
\end{equation}
This point evades constraints from LEP~\cite{Braibant:2003px,Heister:2002hp,Abdallah:2003xe,Achard:2003ge,Abbiendi:2002mp}, has reasonable coupling sizes, scalar potential stability, and is within the $2\sigma$ limits on \hgg~\cite{Aaboud:2018xdt}.  Full cancellation of the NP diphoton decay contributions is possible --- however to avoid tuning parameters we are content with values that respect current bounds.
This point also saturates ATLAS~\cite{Aad:2020plj} upper limit on \zg. We take this as a representative of the relatively complicated models required to generate a large anomaly in \zg~while being consistent with other constraints.

\subsection{Four-Lepton Sensitivity}

Here we calculate the number of \hl~events required to probe the above benchmark point in the singlet-triplet model using the analysis described in section~\ref{sec:4l}.
%
% Four lepton analysis \cite{Chen:2014gka, Chen:2015iha, Chen:2016ofc, Chen:2015iha} is used due to the extra sensitivity provided by the large number of analyzable kinematic factors. 
%
%Using this extremized point, 
From the loop diagrams in Fig.\,\ref{fig:BSMContros}, we can determine the contributions to the effective operators $A^{\gamma\gamma}_{2}$ and $A^{Z\gamma}_{2}$:
\begin{equation}\label{eqn:STOppContros}
\begin{split}
\frac{A^{\gamma\gamma}_{2}({\rm NP})}{A^{\gamma\gamma}_{2}({\rm SM})} =  \frac{-0.00122}{-0.008} = 0.1525
\\
\frac{A^{Z\gamma}_{2}({\rm NP})}{A^{Z\gamma}_{2}({\rm SM})} = \frac{-0.00855}{-0.014} = 0.6107
\end{split}
\end{equation} 
The modification of the branching ratio (or event rate) is given $(1 + A({\rm NP})/A({\rm SM}))^2$. In terms of $\kappa_\gamma$ and $\kappa_{Z\gamma}$:
\begin{equation}\label{eqn:STOppContros2}
\begin{split}
\kappa_\gamma =  1.1525 \,\,\,
\\
\kappa_{Z\gamma} = 1.6107 \, .
\end{split}
\end{equation} 

%%
%\begin{equation}\label{eqn:STOppContros}
%\begin{split}
%\left| \frac{A^{\gamma\gamma}_{2BSM}}{A^{\gamma\gamma}_{2SM}}\right| = \left| \frac{-0.00122}{-0.008}\right| = 0.1525
%\\
%\left| \frac{A^{Z\gamma}_{2BSM}}{A^{Z\gamma}_{2SM}}\right| = \left| \frac{-0.00855}{-0.014}\right| = 0.6107
%\end{split}
%\end{equation} 
%%

\begin{figure*}[tb]
\centering
\begin{minipage}[c]{\textwidth}
\includegraphics[width=0.40\textwidth]{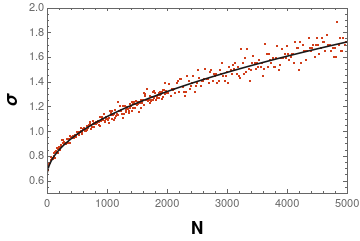}
\includegraphics[width=0.40\textwidth]{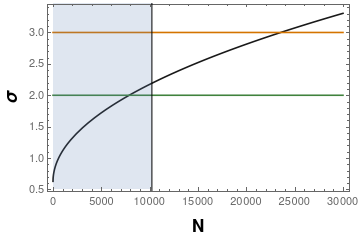}
\end{minipage}
\hfill
\caption{Number of Higgs to four lepton events $N$ required to distinguish the SM from a large $Z\gamma$ model benchmark at a given significance $\sigma$. The left figure shows the curve fitted to the average statistical significance of pseudo-experiments featuring a given number of events (the red points). The right figure shows an extrapolation of this data with lines at 2 and 3 $\sigma$ significance. The blue shaded region shows the expected reach of the HL LHC. 
}
\label{fig:myLGResults}
\end{figure*}

 Applying the $4\ell$ analysis, we obtain the results shown in Fig.\,\ref{fig:myLGResults}. Specifically, we generated 1 million MC sample $pp \rightarrow h \rightarrow 4l$ events for both the SM scenario and our singlet-triplet benchmark, and the liklihood was constructed using multiple pseudo-experiments with varying number of events $N$. 
 %for the 14 TeV LHC using MadGraph5\texttt{\textunderscore}aMC$@$NLO. 
 %The relevant kinematic information for four-lepton analysis was gathered from these simulated events and the procedure in Sec.\,\ref{sec:4l} was used to determine the discriminating power between scenarios as a function of the number of signal events $N$. 
 In Fig.\,\ref{fig:myLGResults}, the points in the left figure show the discrimination power between to the SM and NP models presented in terms of Gaussian $\sigma$ as a function of $N$, the number of events. We also impose a fit curve of the form $x_0 + x_1\sqrt{N}$ as $\sqrt{N}$ growth is expected for large $N$ with large signal to background, as is the case in \hl.   

%used to generate two likelihoods for a given set of $N$ events: one assuming SM physics, the other assuming NP contributions. Using Eq.\,\ref{eqn:testStat}, a test statistic was generated. This process, taking one set of $N$ events from SM simulated events and another set of $N$ events from the NP simulated events then determining two likelihoods from each (one where SM physics is assumed and one where NP physics is assumed) a test statistic is calculated a single pseudo-experiment. Running multiple pseudo-experiments for $N$ events allows for the building up of two test statistic distributions. The statistical separation of these distributions is calculated using Eq.\,\ref{eqn:testStatComp} giving us a confidence value for distinguishing the scenarios given $N$ events. In Fig.\,\ref{fig:myLGResults}, the points in the left figure show the $\sigma$ value calculated from the pseudo-experiments for $N$ signal events.

%Explicitly, for $N \leq 1000$, $1000$ event pseudo-experiments were run and pseudo-experiments were run at 10 event intervals. As values became larger, the number of pseudo-experiments gradually declined (to avoid running out of simulated events) to a minimum of 100 and for $N > 2000$ pseudo-experiments were run at 20 event internals. Once these points were generated upto $N = 10,000$, a curve of the form $a + b\sqrt{N}$ was fit to the data --- giving us our line in the left figure.   

The curve from the left figure is extrapolated in the right part of Fig.\,\ref{fig:myLGResults} in order to estimate the required number of events to distinguish these two scenarios. With $\sim 8000$ events, we can rule out the parameter point at $2\sigma$ and, more generally, we can begin to probe the large $Z\gamma$ model parameter space. Using $139\,\, \mathrm{fb}^{-1}$ of data, ATLAS has measured the fiducial cross section of the $H \rightarrow ZZ^{*}\rightarrow 4l$ to be $\sigma_{fid} = 3.28 \pm 0.32$ fb in good agreement with the SM prediction $\sigma_{fid} = 3.41 \pm 0.18$ fb \cite{ATLAS:2020wny}. With the high luminosity (HL) LHC expected to collect upwards of $3000 \,\, \mathrm{fb}^{-1}$ over its lifetime~\cite{Schmidt_2016}, the LHC should be able to begin seriously probing the large $Z\gamma$ parameter space by the end of its run.

At a future high energy hadron collider the prospects improve further. At $100$ TeV center of mass, the expected gluon fusion production cross section is $\sigma = 808.23^{+44.53}_{-56.95}$ pb~\cite{Dulat:2018rbf} with additional contributions of $\sim 10$\% total from vector boson fusion and associated production \cite{QCDEW2015}. Taking the $h \rightarrow 4l$ branching ratio to be $2.796 \times 10^{-4}$~\cite{deFlorian:2016spz} and assuming an integrated luminosity of $10 \,\mathrm{ab}^{-1}$ gives $\sim 2.5$ million signal events --- enough to potentially probe the relevant parameter space of this model. With this level of precision, other effects not considered here such as backgrounds, detector effects, and higher order corrections may become important.
 
\section{Supersymmetric Models}
\label{sec:SUSY}

%With the viability of probing large $Z\gamma$ models established, we turn our attention to the (previously) highly anticipated, amazingly well motivated, and all-time number one heart-breaker of cynical physicists: the ever elusive and yet always appealing supersymmetry. 
We now turn to study supersymmetric models~\cite{Martin:1997ns} to see if large effects in \zg~can appear in well motivated models. Using the analysis techniques of Sec.~\ref{sec:LargeZgamma}, we examine scalar top partners which, in the MSSM, generically have the largest coupling to the Higgs of new supersymmetric states. 
%--- its scalar particles (in this case supersymmetric top partners: ``stops") that give loop contributions (as shown in Fig. \ref{fig:BSMContros}) to our effective $A^{\gamma\gamma}_{2}$ and $A^{Z\gamma}_2$ operators. 
%
Unsurprisingly, the additional restrictions coming from the extra structure present in SUSY models leads to much smaller allowed contributions to \zg~than with the general multiplet scenarios. 
%The large $Z\gamma$ parameter space, in particular, does not exist within this class of models: 
In addition, for SUSY models we find that the deviations in the branching ratio of \zg~are always smaller than the deviations in \hgg.
%contributions to $Z\gamma$ decay are smaller (when normalized to the size of the SM decay) than the contributions to diphoton decay. 

In addition to considering stops in the MSSM, we also consider F-stops, scalar top partners in folded SUSY models~\cite{Burdman:2006tz}. This class of models stabilizes the weak scale against radiative corrections to around $5\,\tev$. 
%Just like in SUSY, quadratic divergences are cancelled by opposite spin partners; however, unlike SUSY, the gauge quantum numbers of new supersymmetric particles are not identical to their standard model partners. Specifically, 
The scalar partners in these models are not charged under SU$(3)_c$, but have the same electroweak quantum numbers as stops allowing them to couple to the Higgs in the same way. This results in F-SUSY contributions to the Higgs decays to two photons and $Z\gamma$, but no modification to the $gg\rightarrow h$ amplitude (although there is a new decay to hidden gluons, $h \rightarrow g_h g_h$). There are many versions of F-SUSY to chose from; here we take the simplest approach and imagine a scenario identical to the MSSM stop sector sans stop-gluon couplings~\cite{Fan:2014txa}.
We examine the current restrictions on the stop parameter space for both models and then complete a four lepton analysis to estimate the required number of Higgs events to probe these scenarios.

The (F)-stop parameter space is dominantly controlled by three parameters: the two mass eigenvalues $m_1$, $m_2$, and the mixing angle between the gauge and mass eigenstates, $\theta_t$ (we use the conventions in~\cite{Batell:2015koa}). There is some weak dependence on $\tan\beta$ --- the couplings change less than $10\%$ over any possible $\beta$ value and, if we restrict this parameter to its normally accepted values, $3 \leq \tan{\beta} \leq 50$, the modification becomes $\cO (1\%)$. As such, we take $\tan{\beta} = 10$ for simplicity.

\begin{figure*}[tb]
\centering
\begin{minipage}[c]{\textwidth}
\includegraphics[width=0.40\textwidth]{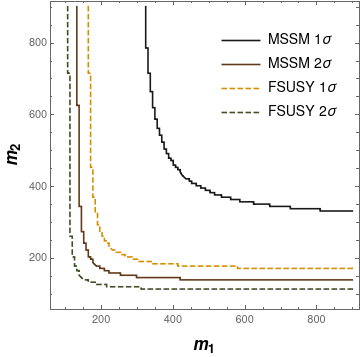}
\includegraphics[width=0.40\textwidth]{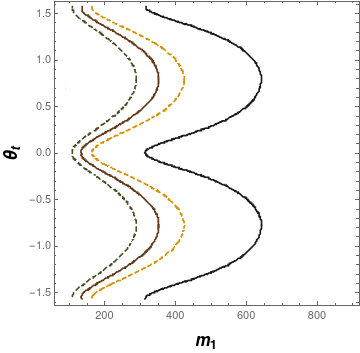}
\end{minipage}
\hfill
\caption{Exclusion bounds in stop parameter space due to \hgg~and gluon fusion measurements: the solid rust (black) lines represent the 2 (1) sigma exclusion bounds for  stops and the dashed green (yellow) lines represent the 2 (1) sigma exclusion bounds for F-stops. The left figure is in the stop mass 1, stop mass 2 plane with zero mixing. The right figure is in the stop mass 1, stop mixing angle plane, with stop mass 2 held at 1 TeV.} 
\label{fig:myDPResults}
\end{figure*}

Using standard formulas in the literature we can compute the contributions of stops to $gg\rightarrow h$ and \hgg~and apply the ATLAS constraints shown in Fig.~\ref{fig:aDPResults}. We ignore direct bounds on (F)-stops in order to keep our constraints model independent. As this is a three dimensional parameter space, we present our results in Fig.~\ref{fig:myDPResults} in two dimensional slices of the two mass eigenstates with the mixing angle fixed at $\theta_t = 0$ (left) and the lighter mass eigenstate $m_1$ vs.~the mixing angle $\theta_t$ with $m_2 = 1$ TeV. We note that these results are leading order, and higher order corrections place an uncertainty on these bounds of $\mathcal{O}(10\%)$. 
This is an update of the analysis in~\cite{Fan:2014txa} using significantly more data. Scanning the parameter space indicates that MSSM stops below $\sim 140 \,\gev$ are essentially excluded. Additionally, both stops having a mass under $\sim 200 \, \gev$ is excluded. In general, the weakest parameter space constraints occur when either mixing is maximized or the stops have a very large mass splitting and are effectively decoupled. 

%Working with the stop sector outlined above, we can start to place restrictions on the MSSM parameter space. Our four dimensional parameter space can be reduced slightly by noting that the only place $\tan{\beta}$ enters process is through our NP stop couplings (Eq. (\ref{eqn:hstopcouplings})). The value of these couplings changes less than $10\%$ over any possible $\beta$ value and, if we restrict this parameter to its normall accepted values, $3 \leq \tan{\beta} \leq 50$, the modification becomes $\cO (1\%)$. As such, we take $\tan{\beta} = 10$ for simplicity.

%Ultimately, the limits set by the ATLAS diphoton data (Fig. \ref{fig:aDPResults}) lead to the exclusion bounds in stop parameter space outlined in Fig. \ref{fig:myDPResults}. 

We can now compute the contribution to the $CP$-even form factor $A_2$ from Eq.~(\ref{eqn:tensorStruc}).\footnote{$CP$ violation can exist in the MSSM but the constraints are strong~\cite{Falk:1995fk}.} By scanning over the $m_1, m_2$, and $\theta_t$ parameter space and subjecting the results to the previously mentioned ATLAS constraints, the maximal modification of said form factor (and thus matrix element) was found to be: 
\begin{equation}\label{eqn:MSSMOppControEst}
\begin{split}
\left| \frac{A^{\gamma\gamma}_{2}(\rm{stop})}{A^{\gamma\gamma}_{2}(\rm{SM})}\right| \lesssim 9\%
\\
\left| \frac{A^{Z\gamma}_{2}(\rm{stop})}{A^{Z\gamma}_{2}(\rm{SM})}\right| \lesssim 4\%.
\end{split}
\end{equation} 
This is beyond the precision expected on the coupling $\kappa_{Z\gamma}$ from the HL-LHC of $\mathcal{O}(10$\%)~\cite{HLLHC}, but deviations from the SM in this channel may be visible at a 100 TeV collider with an expected precision of $\mathcal{O}(1$\%)~\cite{Contino:2016spe}, but this would require reducing the uncertainty in the SM prediction which is currently $\mathcal{O}(5$\%) level~\cite{deFlorian:2016spz}. If this is the theory of nature, we expect to see deviation from the SM in \hgg\footnote{Note that these deviations are significantly larger than the uncertainty on the SM prediction which is of $\mathcal{O}(1$\%)~\cite{deFlorian:2016spz}.} long before we see one in \zg.

%\subsection{Folded SUSY}
%\label{sec:FSUSY}

%\begin{figure*}[tb]
%\centering
%\begin{minipage}[c]{\textwidth}
%\includegraphics[width=0.40\textwidth]{diphotonResFSUSYMM.png}
%\includegraphics[width=0.40\textwidth]{diphotonResFSUSYMA.png}
%\end{minipage}
%\hfill
%\caption{Exclusion bounds in Folded-SUSY stop parameter space due to diphoton measurements; the orange (blue) lines represent the 2 (1) sigma exclusion bounds respectively. The left figure is in the stop mass 1, stop mass 2 plane with zero mixing. The right figure is in the stop mass 1, stop mixing angle plane, with stop mass 2 held at 1 TeV.
%}
%\label{fig:myDPFSUSYResults}
%\end{figure*}

 For F-stops, the procedure is the same, but we can ignore constraints from gluon fusion production of the Higgs. 
 This makes measurements less constricting leading to the results presented with the dashed lines in Fig.~\ref{fig:myDPResults}.
The notably weaker constraints lead to stops only below $\sim 100 \,\gev$ being universally excluded and requiring at least one stop above $\sim 150\,\gev$.
%Perhaps most notable of all is how much the $1\sigma$ bounds weaken --- from both stops being around $400\, \gev$ to $200\, \gev$. 
Other than the strength of the bounds, the overall takeaways remain the same as with MSSM stops: the weakest constraints on the lighter stop occur when either mixing angle has values of $\theta_t = 0, \pm \frac{\pi}{2}$ or the stops have a very large mass splitting.
The $A_2$ values in the F-stop allows greater deviations:
\begin{equation}\label{eqn:FSUSYOppControEst}
\begin{split}
\left| \frac{A^{\gamma\gamma}_{2}(\mbox{F-stop})}{A^{\gamma\gamma}_{2}(\rm{SM})}\right| \lesssim 14\%
\\
\left| \frac{A^{Z\gamma}_{2}(\mbox{F-stop})}{A^{Z\gamma}_{2}(\rm{SM})}\right| \lesssim 9\%.
\end{split}
\end{equation}  
but we see that the deviations in \zg~are much smaller than the singlet triplet model shown in Eq.~\eqref{eqn:STOppContros}. This is just on the edge of the sensitivity of the HL-LHC in on-shell \zg~\cite{HLLHC}, but deviations would first be seen in \hgg. 

\subsection{Four-Lepton Analysis}

The four-lepton analysis of the SUSY models follows the one in Sec.\,\ref{sec:LargeZgamma}. 
%MadGraph5\texttt{\textunderscore}aMC$@$NLO was used to generate MC events for MSSM and FSUSY parameter points. 
%Restrictions on the parameter space from diphoton decay bounds (Fig.\,\ref{fig:myDPResults}) were imposed and points that had relatively large BSM contributions to the operators $A^{\gamma\gamma}_{2}$ and $A^{Z\gamma}_{2}$ were selected. Just as before, this was done to check the viability of even beginning to search the four-lepton decay channel for these signals at the LHC and beyond.
 For stops, we use the following benchmark point:
\begin{equation}\label{eqn:MSSMParams}
\begin{split}
\bullet &\,\, m_{1} = 177.8\,\, \mathrm{GeV}
\\
\bullet &\,\, m_{2} = 184.0\,\, \mathrm{GeV}
\\
\bullet &\,\, \theta_t = -1.435 .
\end{split}
\end{equation}
%
%with $m_{\tilde{t}_1}$ and $m_{\tilde{t}_2}$ being the masses of the two stops and $\theta_t$ being their mixing angle. 
which gives form factors
\begin{equation}\label{eqn:MSSMOppContros}
\begin{split}
 \frac{A^{\gamma\gamma}_{2}(\mbox{stop})}{A^{\gamma\gamma}_{2}(\rm{SM})} = \frac{0.00072}{-0.008} = -0.09
\\
 \frac{A^{Z\gamma}_{2}(\mbox{stop})}{A^{Z\gamma}_{2}(\rm{SM})} =  \frac{0.00028}{-0.014} = -0.02
\\
\frac{A^{gg}_{2}(\mbox{stop})}{A^{gg}_{2}(\mbox{SM})} = \frac{0.00075}{0.0127} = 0.0593
\end{split}
\end{equation}
where we have introduced the gluon form factor $A_2^{gg}$ analogous to those of the electroweak gauge bosons. Similarly, the point chosen for F-stops has parameters,
\begin{equation}\label{eqn:FSUSYParams}
\begin{split}
\bullet &\,\, m_{1} = 126.2\,\, \mathrm{GeV}
\\
\bullet &\,\, m_{2} = 180.6\,\, \mathrm{GeV}
\\
\bullet &\,\, \theta_t = -1.285.
\end{split}
\end{equation}
which gives:
\begin{equation}\label{eqn:FSUSYOppContros}
\begin{split}
 \frac{A^{\gamma\gamma}_{2}(\mbox{F-stop})}{A^{\gamma\gamma}_{2}(\rm{SM})}= \frac{0.00115}{-0.008}= - 0.1437
\\
 \frac{A^{Z\gamma}_{2}(\mbox{F-stop})}{A^{Z\gamma}_{2}(\rm{SM})} = \frac{0.00082}{-0.014}= -0.0586.
\end{split}
\end{equation}  

\begin{figure*}[tb]
\centering
\begin{minipage}[c]{\textwidth}
\includegraphics[width=0.40\textwidth]{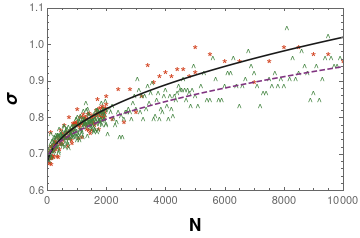}
\includegraphics[width=0.40\textwidth]{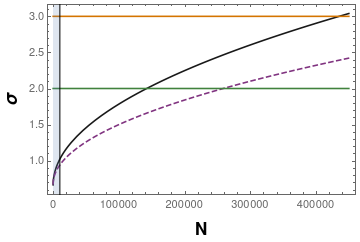}
\end{minipage}
\hfill
\caption{Number of Higgs to four lepton events $N$ required to distinguish the SM from the SUSY model benchmarks at a given significance $\sigma$. The left figure shows the curve fitted to the average statistical significance of pseudo-experiments featuring a given number of events. The dashed violet line (fit to the green carets) is the stop scenario and the solid black line (fit to the red asterisks) is the F-stop scenario. The right figure shows an extrapolation of this data with lines at 2 and 3 $\sigma$ significance. The thin blue band on the left is the reach of the HL-LHC.
}
\label{fig:mySUSYResults}
\end{figure*}

As in the large $Z\gamma$ case, these coupling modifications were used in generating MC events that were then processed into likelihoods, test statistic distributions, and, finally, statistical confidence as a function of number of signal events $\sigma(N)$. The results for both stop and F-stop scenarios are shown in Fig.\,\ref{fig:mySUSYResults}. Unsurprisingly, the much smaller impact of the intermediate scalar particles within the loop decays makes both the stop and F-stop much more difficult to probe than the large $Z\gamma$ models. With the HL-LHC's integrated luminosity topping out around $3000 \,\, \mathrm{fb}^{-1}$ after its final run~\cite{Schmidt_2016}, the required $\sim 150$ thousand events to be sensitive F-stops or $\sim 270$ thousand events required to explore stops at $2\sigma$ are not reachable with the LHC. These numbers are, however, potentially in reach of a future $100$ TeV collider --- having an integrated luminosity on the order $10 \,\mathrm{ab}^{-1}$ gives $\sim 2.5$ million signal events which can potentially probe the SUSY parameter space. 

\section{Conclusion}\label{sec:Conclusion}

We have explored the possibility of new physics models with significant enhancements to the \zg~decay. We have demonstrated that simple models cannot give rise to large enhancements while still being consistent with other data, particularly measurements of \hgg. Models with multiple multiplets featuring auspicious cancellations could produce such a signal. For models that do feature large contributions to \zg, such as a model with a singlet and a triplet explored in section~\ref{sec:singletTriplet}, kinematic analysis of \hl~will be able to probe these models with the data from the high-luminosity LHC. 

We also explored more motivated models that can solve the hierarchy problem, focusing in particular on stop contributions in supersymmetry and colourless top partners (F-stops) in folded SUSY. We used current Higgs data to constrain the parameter space of those states updating the analysis of~\cite{Fan:2014txa}, and determined these models cannot give measurable deviations to \zg. Kinematic analysis of \hl~can only have sensitivity to these models with significantly more data than the LHC will have although this may be possible with a future 100 TeV hadron collider. 

Although this work focused exclusively on the new physics contributions of scalars, an analagous analysis with fermionic multiplets would not change the qualitative conclusions: the loop factors and Higgs couplings are slightly different, but the photon and $Z$ couplings still arise from the covariant derivative and models with large contributions to \zg~can only be produced through fortuitous cancellation by multiple new physics multiplets. $CP$ violating couplings were also not examined in this work, but the constraints on Higgs couplings due to EDM measurements~\cite{Brod:2013cka} indicate that such couplings would have to be tiny --- much too small to achieve the large $Z\gamma$ contributions that we examined.    

What this ultimately means is that, in general, the \zg~channel is unlikely to be a place where new physics is discovered: simple or motivated models tend to lead to contributions smaller than the much more sensitive diphoton channel. As such, a discovery of a significant NP contribution to this channel is a strong indication that interference effects are present in the NP sector and points towards non-minimal models like those presented in this paper. Finally, the discriminating power of a $100$ TeV collider, particularly in \hl, is undeniable: where the LHC cannot even probe the most favourable parameter points of the MSSM, a future collider will collect significantly more events, potentially giving us a much clearer handle on the nature of our microscopic world.

%----------------------------------------------------------------
% Acknowledgements
%----------------------------------------------------------------
\subsection*{Acknowledgements}
The authors thank Yi Chen, Kevin Earl, and Dylan Linthorne for helpful discussions. P.A.S~is supported in part by the Ontario Graduate Scholarship (OGS). D.S.~is supported in part by the Natural Sciences and Engineering Research Council of Canada (NSERC). The work of R.V.M.~has been partially supported by the Ministry of Science and Innovation under grant number FPA2016-78220-C3-1-P (FEDER), SRA under grant PID2019-106087GB-C22 (10.13039/501100011033), and by the Junta de Andaluc\'{i}a grants FQM 101, and A-FQM-211-UGR18 and P18-FR-4314 (FEDER).

%%%%%%%%%%%%%%%%%%%%%%%%%
%%%%%%%%%%%%%%%%%%%%%%%%%
%%%%%%%%%%%%%%%%%%%%%%%%%
%%%%%%%%%%%%%%%%%%%%%%%%%
%%%%%%%%%%%%%%%%%%%%%%%%%

\appendix

\section{Loop integral definitions}\label{app:loops}

Here, for completeness, we present the integrated loop functions $l_{\gamma\gamma}$ and $l_{Z\gamma}$ for scalar NP particles of mass $m$ mentioned in Eq.~(\ref{eqn:LZGOppForm}). For further details, please see~\cite{Djouadi:2005gj,Carena:2012xa}.

The diphoton expression:
\begin{equation}
l_{\gamma\gamma}[m]=\left(\frac{v}{4 \pi ^2 m_h^4}\right) \left(m_h^2-4 m^2 \tan ^{-1}\left(\frac{m_h}{\sqrt{4 m^2-m_h^2}}\right)^2\right)
\end{equation}
And the $Z\gamma$ equation:
\small
\begin{widetext}
%\[
\begin{equation}
\begin{split}
l_{Z\gamma}[m]=&\frac{v}{4 \pi ^2 m_h \left(m_h^2-m_Z^2\right)^2} \Bigg\{m_h^3-2 m_Z^2 \sqrt{4 m^2-m_h^2} \mathrm{tan}^{-1}\left[\frac{m_h}{\sqrt{4 m^2-m_h^2}}\right]
\\
&-m_h \left[2 m^2 \left[2 \tan ^{-1}\left[\frac{m_h}{\sqrt{4 m^2-m_h^2}}\right]^2-2 \tan ^{-1}\left[\frac{m_Z}{\sqrt{4 m^2-m_Z^2}}\right]^2\right]-2 m_Z \sqrt{4 m^2-m_Z^2} \tan ^{-1}\left[\frac{m_Z}{\sqrt{4 m^2-m_Z^2}}\right]+m_Z^2\right]\Bigg\}
\end{split}
\end{equation}

%\]
\end{widetext}

\normalsize

For the $Z\gamma$ expression, the case presented here describes loops with only one type of new particle; loops where the legs are different particles lead to longer expressions we do not present here. In the case of the singlet-triplet model presented in Sec.~\ref{sec:singletTriplet}, this expression holds --- however for the SUSY models in Sec.~\ref{sec:SUSY} the more general expression is required due to the possibility of the presence of both stops.
\bibliographystyle{apsrev4-2}
\bibliography{refs}

\begin{thebibliography}{90}%
\makeatletter
\providecommand \@ifxundefined [1]{%
 \@ifx{#1\undefined}
}%
\providecommand \@ifnum [1]{%
 \ifnum #1\expandafter \@firstoftwo
 \else \expandafter \@secondoftwo
 \fi
}%
\providecommand \@ifx [1]{%
 \ifx #1\expandafter \@firstoftwo
 \else \expandafter \@secondoftwo
 \fi
}%
\providecommand \natexlab [1]{#1}%
\providecommand \enquote  [1]{``#1''}%
\providecommand \bibnamefont  [1]{#1}%
\providecommand \bibfnamefont [1]{#1}%
\providecommand \citenamefont [1]{#1}%
\providecommand \href@noop [0]{\@secondoftwo}%
\providecommand \href [0]{\begingroup \@sanitize@url \@href}%
\providecommand \@href[1]{\@@startlink{#1}\@@href}%
\providecommand \@@href[1]{\endgroup#1\@@endlink}%
\providecommand \@sanitize@url [0]{\catcode `\\12\catcode `\$12\catcode
  `\&12\catcode `\#12\catcode `\^12\catcode `\_12\catcode `\%12\relax}%
\providecommand \@@startlink[1]{}%
\providecommand \@@endlink[0]{}%
\providecommand \url  [0]{\begingroup\@sanitize@url \@url }%
\providecommand \@url [1]{\endgroup\@href {#1}{\urlprefix }}%
\providecommand \urlprefix  [0]{URL }%
\providecommand \Eprint [0]{\href }%
\providecommand \doibase [0]{https://doi.org/}%
\providecommand \selectlanguage [0]{\@gobble}%
\providecommand \bibinfo  [0]{\@secondoftwo}%
\providecommand \bibfield  [0]{\@secondoftwo}%
\providecommand \translation [1]{[#1]}%
\providecommand \BibitemOpen [0]{}%
\providecommand \bibitemStop [0]{}%
\providecommand \bibitemNoStop [0]{.\EOS\space}%
\providecommand \EOS [0]{\spacefactor3000\relax}%
\providecommand \BibitemShut  [1]{\csname bibitem#1\endcsname}%
\let\auto@bib@innerbib\@empty
%</preamble>
\bibitem [{\citenamefont {Aaboud}\ \emph {et~al.}(2019)\citenamefont {Aaboud}
  \emph {et~al.}}]{Aaboud:2018jqu}%
  \BibitemOpen
  \bibfield  {author} {\bibinfo {author} {\bibfnamefont {M.}~\bibnamefont
  {Aaboud}} \emph {et~al.} (\bibinfo {collaboration} {ATLAS}),\ }\href
  {https://doi.org/10.1016/j.physletb.2018.11.064} {\bibfield  {journal}
  {\bibinfo  {journal} {Phys. Lett. B}\ }\textbf {\bibinfo {volume} {789}},\
  \bibinfo {pages} {508} (\bibinfo {year} {2019})},\ \Eprint
  {https://arxiv.org/abs/1808.09054} {arXiv:1808.09054 [hep-ex]} \BibitemShut
  {NoStop}%
\bibitem [{\citenamefont {Aad}\ \emph {et~al.}(2019)\citenamefont {Aad} \emph
  {et~al.}}]{Aad:2019lpq}%
  \BibitemOpen
  \bibfield  {author} {\bibinfo {author} {\bibfnamefont {G.}~\bibnamefont
  {Aad}} \emph {et~al.} (\bibinfo {collaboration} {ATLAS}),\ }\href
  {https://doi.org/10.1016/j.physletb.2019.134949} {\bibfield  {journal}
  {\bibinfo  {journal} {Phys. Lett. B}\ }\textbf {\bibinfo {volume} {798}},\
  \bibinfo {pages} {134949} (\bibinfo {year} {2019})},\ \Eprint
  {https://arxiv.org/abs/1903.10052} {arXiv:1903.10052 [hep-ex]} \BibitemShut
  {NoStop}%
\bibitem [{\citenamefont {Sirunyan}\ \emph
  {et~al.}(2019{\natexlab{a}})\citenamefont {Sirunyan} \emph
  {et~al.}}]{Sirunyan:2018egh}%
  \BibitemOpen
  \bibfield  {author} {\bibinfo {author} {\bibfnamefont {A.~M.}\ \bibnamefont
  {Sirunyan}} \emph {et~al.} (\bibinfo {collaboration} {CMS}),\ }\href
  {https://doi.org/10.1016/j.physletb.2018.12.073} {\bibfield  {journal}
  {\bibinfo  {journal} {Phys. Lett. B}\ }\textbf {\bibinfo {volume} {791}},\
  \bibinfo {pages} {96} (\bibinfo {year} {2019}{\natexlab{a}})},\ \Eprint
  {https://arxiv.org/abs/1806.05246} {arXiv:1806.05246 [hep-ex]} \BibitemShut
  {NoStop}%
\bibitem [{\citenamefont {Sirunyan}\ \emph {et~al.}(2020)\citenamefont
  {Sirunyan} \emph {et~al.}}]{Sirunyan:2020tzo}%
  \BibitemOpen
  \bibfield  {author} {\bibinfo {author} {\bibfnamefont {A.~M.}\ \bibnamefont
  {Sirunyan}} \emph {et~al.} (\bibinfo {collaboration} {CMS}),\ }\href@noop {}
  {\  (\bibinfo {year} {2020})},\ \Eprint {https://arxiv.org/abs/2007.01984}
  {arXiv:2007.01984 [hep-ex]} \BibitemShut {NoStop}%
\bibitem [{\citenamefont {Sirunyan}\ \emph
  {et~al.}(2019{\natexlab{b}})\citenamefont {Sirunyan} \emph
  {et~al.}}]{Sirunyan:2018sgc}%
  \BibitemOpen
  \bibfield  {author} {\bibinfo {author} {\bibfnamefont {A.~M.}\ \bibnamefont
  {Sirunyan}} \emph {et~al.} (\bibinfo {collaboration} {CMS}),\ }\href
  {https://doi.org/10.1016/j.physletb.2019.03.059} {\bibfield  {journal}
  {\bibinfo  {journal} {Phys. Lett. B}\ }\textbf {\bibinfo {volume} {792}},\
  \bibinfo {pages} {369} (\bibinfo {year} {2019}{\natexlab{b}})},\ \Eprint
  {https://arxiv.org/abs/1812.06504} {arXiv:1812.06504 [hep-ex]} \BibitemShut
  {NoStop}%
\bibitem [{\citenamefont {Aad}\ \emph {et~al.}(2013)\citenamefont {Aad} \emph
  {et~al.}}]{Aad:2013wqa}%
  \BibitemOpen
  \bibfield  {author} {\bibinfo {author} {\bibfnamefont {G.}~\bibnamefont
  {Aad}} \emph {et~al.} (\bibinfo {collaboration} {ATLAS}),\ }\href
  {https://doi.org/10.1016/j.physletb.2014.05.011} {\bibfield  {journal}
  {\bibinfo  {journal} {Phys. Lett. B}\ }\textbf {\bibinfo {volume} {726}},\
  \bibinfo {pages} {88} (\bibinfo {year} {2013})},\ \bibinfo {note} {[Erratum:
  Phys.Lett.B 734, 406--406 (2014)]},\ \Eprint
  {https://arxiv.org/abs/1307.1427} {arXiv:1307.1427 [hep-ex]} \BibitemShut
  {NoStop}%
\bibitem [{\citenamefont {Aaboud}\ \emph {et~al.}(2018)\citenamefont {Aaboud}
  \emph {et~al.}}]{Aaboud:2018xdt}%
  \BibitemOpen
  \bibfield  {author} {\bibinfo {author} {\bibfnamefont {M.}~\bibnamefont
  {Aaboud}} \emph {et~al.} (\bibinfo {collaboration} {ATLAS}),\ }\href
  {https://doi.org/10.1103/PhysRevD.98.052005} {\bibfield  {journal} {\bibinfo
  {journal} {Phys. Rev.}\ }\textbf {\bibinfo {volume} {D98}},\ \bibinfo {pages}
  {052005} (\bibinfo {year} {2018})},\ \Eprint
  {https://arxiv.org/abs/1802.04146} {arXiv:1802.04146 [hep-ex]} \BibitemShut
  {NoStop}%
%%CITATION = ARXIV:1802.04146;%%
\bibitem [{\citenamefont {Aad}\ \emph {et~al.}(2020{\natexlab{a}})\citenamefont
  {Aad} \emph {et~al.}}]{Aad:2020plj}%
  \BibitemOpen
  \bibfield  {author} {\bibinfo {author} {\bibfnamefont {G.}~\bibnamefont
  {Aad}} \emph {et~al.} (\bibinfo {collaboration} {ATLAS}),\ }\href
  {https://doi.org/10.1016/j.physletb.2020.135754} {\bibfield  {journal}
  {\bibinfo  {journal} {Phys. Lett. B}\ }\textbf {\bibinfo {volume} {809}},\
  \bibinfo {pages} {135754} (\bibinfo {year} {2020}{\natexlab{a}})},\ \Eprint
  {https://arxiv.org/abs/2005.05382} {arXiv:2005.05382 [hep-ex]} \BibitemShut
  {NoStop}%
\bibitem [{\citenamefont {Sirunyan}\ \emph {et~al.}(2018)\citenamefont
  {Sirunyan} \emph {et~al.}}]{Sirunyan:2018tbk}%
  \BibitemOpen
  \bibfield  {author} {\bibinfo {author} {\bibfnamefont {A.~M.}\ \bibnamefont
  {Sirunyan}} \emph {et~al.} (\bibinfo {collaboration} {CMS}),\ }\href
  {https://doi.org/10.1007/JHEP11(2018)152} {\bibfield  {journal} {\bibinfo
  {journal} {JHEP}\ }\textbf {\bibinfo {volume} {11}},\ \bibinfo {pages}
  {152}},\ \Eprint {https://arxiv.org/abs/1806.05996} {arXiv:1806.05996
  [hep-ex]} \BibitemShut {NoStop}%
\bibitem [{\citenamefont {Cahn}\ \emph {et~al.}(1979)\citenamefont {Cahn},
  \citenamefont {Chanowitz},\ and\ \citenamefont {Fleishon}}]{Cahn:1978nz}%
  \BibitemOpen
  \bibfield  {author} {\bibinfo {author} {\bibfnamefont {R.}~\bibnamefont
  {Cahn}}, \bibinfo {author} {\bibfnamefont {M.~S.}\ \bibnamefont
  {Chanowitz}},\ and\ \bibinfo {author} {\bibfnamefont {N.}~\bibnamefont
  {Fleishon}},\ }\href {https://doi.org/10.1016/0370-2693(79)90438-6}
  {\bibfield  {journal} {\bibinfo  {journal} {Phys. Lett. B}\ }\textbf
  {\bibinfo {volume} {82}},\ \bibinfo {pages} {113} (\bibinfo {year}
  {1979})}\BibitemShut {NoStop}%
\bibitem [{\citenamefont {Bergstrom}\ and\ \citenamefont
  {Hulth}(1985)}]{Bergstrom:1985hp}%
  \BibitemOpen
  \bibfield  {author} {\bibinfo {author} {\bibfnamefont {L.}~\bibnamefont
  {Bergstrom}}\ and\ \bibinfo {author} {\bibfnamefont {G.}~\bibnamefont
  {Hulth}},\ }\href {https://doi.org/10.1016/0550-3213(85)90302-5} {\bibfield
  {journal} {\bibinfo  {journal} {Nucl. Phys. B}\ }\textbf {\bibinfo {volume}
  {259}},\ \bibinfo {pages} {137} (\bibinfo {year} {1985})},\ \bibinfo {note}
  {[Erratum: Nucl.Phys.B 276, 744--744 (1986)]}\BibitemShut {NoStop}%
\bibitem [{\citenamefont {Spira}\ \emph {et~al.}(1992)\citenamefont {Spira},
  \citenamefont {Djouadi},\ and\ \citenamefont {Zerwas}}]{Spira:1991tj}%
  \BibitemOpen
  \bibfield  {author} {\bibinfo {author} {\bibfnamefont {M.}~\bibnamefont
  {Spira}}, \bibinfo {author} {\bibfnamefont {A.}~\bibnamefont {Djouadi}},\
  and\ \bibinfo {author} {\bibfnamefont {P.}~\bibnamefont {Zerwas}},\ }\href
  {https://doi.org/10.1016/0370-2693(92)90331-W} {\bibfield  {journal}
  {\bibinfo  {journal} {Phys. Lett. B}\ }\textbf {\bibinfo {volume} {276}},\
  \bibinfo {pages} {350} (\bibinfo {year} {1992})}\BibitemShut {NoStop}%
\bibitem [{\citenamefont {Bonciani}\ \emph {et~al.}(2015)\citenamefont
  {Bonciani}, \citenamefont {Del~Duca}, \citenamefont {Frellesvig},
  \citenamefont {Henn}, \citenamefont {Moriello},\ and\ \citenamefont
  {Smirnov}}]{Bonciani:2015eua}%
  \BibitemOpen
  \bibfield  {author} {\bibinfo {author} {\bibfnamefont {R.}~\bibnamefont
  {Bonciani}}, \bibinfo {author} {\bibfnamefont {V.}~\bibnamefont {Del~Duca}},
  \bibinfo {author} {\bibfnamefont {H.}~\bibnamefont {Frellesvig}}, \bibinfo
  {author} {\bibfnamefont {J.~M.}\ \bibnamefont {Henn}}, \bibinfo {author}
  {\bibfnamefont {F.}~\bibnamefont {Moriello}},\ and\ \bibinfo {author}
  {\bibfnamefont {V.~A.}\ \bibnamefont {Smirnov}},\ }\href
  {https://doi.org/10.1007/JHEP08(2015)108} {\bibfield  {journal} {\bibinfo
  {journal} {JHEP}\ }\textbf {\bibinfo {volume} {08}},\ \bibinfo {pages}
  {108}},\ \Eprint {https://arxiv.org/abs/1505.00567} {arXiv:1505.00567
  [hep-ph]} \BibitemShut {NoStop}%
\bibitem [{\citenamefont {Gehrmann}\ \emph {et~al.}(2015)\citenamefont
  {Gehrmann}, \citenamefont {Guns},\ and\ \citenamefont
  {Kara}}]{Gehrmann:2015dua}%
  \BibitemOpen
  \bibfield  {author} {\bibinfo {author} {\bibfnamefont {T.}~\bibnamefont
  {Gehrmann}}, \bibinfo {author} {\bibfnamefont {S.}~\bibnamefont {Guns}},\
  and\ \bibinfo {author} {\bibfnamefont {D.}~\bibnamefont {Kara}},\ }\href
  {https://doi.org/10.1007/JHEP09(2015)038} {\bibfield  {journal} {\bibinfo
  {journal} {JHEP}\ }\textbf {\bibinfo {volume} {09}},\ \bibinfo {pages}
  {038}},\ \Eprint {https://arxiv.org/abs/1505.00561} {arXiv:1505.00561
  [hep-ph]} \BibitemShut {NoStop}%
\bibitem [{\citenamefont {Hue}\ \emph {et~al.}(2018)\citenamefont {Hue},
  \citenamefont {Arbuzov}, \citenamefont {Hong}, \citenamefont {Nguyen},
  \citenamefont {Si},\ and\ \citenamefont {Long}}]{Hue:2017cph}%
  \BibitemOpen
  \bibfield  {author} {\bibinfo {author} {\bibfnamefont {L.}~\bibnamefont
  {Hue}}, \bibinfo {author} {\bibfnamefont {A.}~\bibnamefont {Arbuzov}},
  \bibinfo {author} {\bibfnamefont {T.}~\bibnamefont {Hong}}, \bibinfo {author}
  {\bibfnamefont {T.~P.}\ \bibnamefont {Nguyen}}, \bibinfo {author}
  {\bibfnamefont {D.}~\bibnamefont {Si}},\ and\ \bibinfo {author}
  {\bibfnamefont {H.}~\bibnamefont {Long}},\ }\href
  {https://doi.org/10.1140/epjc/s10052-018-6349-0} {\bibfield  {journal}
  {\bibinfo  {journal} {Eur. Phys. J. C}\ }\textbf {\bibinfo {volume} {78}},\
  \bibinfo {pages} {885} (\bibinfo {year} {2018})},\ \Eprint
  {https://arxiv.org/abs/1712.05234} {arXiv:1712.05234 [hep-ph]} \BibitemShut
  {NoStop}%
\bibitem [{\citenamefont {Bellazzini}\ and\ \citenamefont
  {Riva}(2018)}]{Bellazzini:2018paj}%
  \BibitemOpen
  \bibfield  {author} {\bibinfo {author} {\bibfnamefont {B.}~\bibnamefont
  {Bellazzini}}\ and\ \bibinfo {author} {\bibfnamefont {F.}~\bibnamefont
  {Riva}},\ }\href {https://doi.org/10.1103/PhysRevD.98.095021} {\bibfield
  {journal} {\bibinfo  {journal} {Phys. Rev. D}\ }\textbf {\bibinfo {volume}
  {98}},\ \bibinfo {pages} {095021} (\bibinfo {year} {2018})},\ \Eprint
  {https://arxiv.org/abs/1806.09640} {arXiv:1806.09640 [hep-ph]} \BibitemShut
  {NoStop}%
\bibitem [{\citenamefont {Dedes}\ \emph {et~al.}(2019)\citenamefont {Dedes},
  \citenamefont {Suxho},\ and\ \citenamefont {Trifyllis}}]{Dedes:2019bew}%
  \BibitemOpen
  \bibfield  {author} {\bibinfo {author} {\bibfnamefont {A.}~\bibnamefont
  {Dedes}}, \bibinfo {author} {\bibfnamefont {K.}~\bibnamefont {Suxho}},\ and\
  \bibinfo {author} {\bibfnamefont {L.}~\bibnamefont {Trifyllis}},\ }\href
  {https://doi.org/10.1007/JHEP06(2019)115} {\bibfield  {journal} {\bibinfo
  {journal} {JHEP}\ }\textbf {\bibinfo {volume} {06}},\ \bibinfo {pages}
  {115}},\ \Eprint {https://arxiv.org/abs/1903.12046} {arXiv:1903.12046
  [hep-ph]} \BibitemShut {NoStop}%
\bibitem [{\citenamefont {Weiler}\ and\ \citenamefont
  {Yuan}(1989)}]{Weiler:1988xn}%
  \BibitemOpen
  \bibfield  {author} {\bibinfo {author} {\bibfnamefont {T.~J.}\ \bibnamefont
  {Weiler}}\ and\ \bibinfo {author} {\bibfnamefont {T.-C.}\ \bibnamefont
  {Yuan}},\ }\href {https://doi.org/10.1016/0550-3213(89)90610-X} {\bibfield
  {journal} {\bibinfo  {journal} {Nucl. Phys. B}\ }\textbf {\bibinfo {volume}
  {318}},\ \bibinfo {pages} {337} (\bibinfo {year} {1989})}\BibitemShut
  {NoStop}%
\bibitem [{\citenamefont {Chiang}\ and\ \citenamefont
  {Yagyu}(2013)}]{Chiang:2012qz}%
  \BibitemOpen
  \bibfield  {author} {\bibinfo {author} {\bibfnamefont {C.-W.}\ \bibnamefont
  {Chiang}}\ and\ \bibinfo {author} {\bibfnamefont {K.}~\bibnamefont {Yagyu}},\
  }\href {https://doi.org/10.1103/PhysRevD.87.033003} {\bibfield  {journal}
  {\bibinfo  {journal} {Phys. Rev. D}\ }\textbf {\bibinfo {volume} {87}},\
  \bibinfo {pages} {033003} (\bibinfo {year} {2013})},\ \Eprint
  {https://arxiv.org/abs/1207.1065} {arXiv:1207.1065 [hep-ph]} \BibitemShut
  {NoStop}%
\bibitem [{\citenamefont {Cao}\ \emph {et~al.}(2013)\citenamefont {Cao},
  \citenamefont {Wu}, \citenamefont {Wu},\ and\ \citenamefont
  {Yang}}]{Cao:2013ur}%
  \BibitemOpen
  \bibfield  {author} {\bibinfo {author} {\bibfnamefont {J.}~\bibnamefont
  {Cao}}, \bibinfo {author} {\bibfnamefont {L.}~\bibnamefont {Wu}}, \bibinfo
  {author} {\bibfnamefont {P.}~\bibnamefont {Wu}},\ and\ \bibinfo {author}
  {\bibfnamefont {J.~M.}\ \bibnamefont {Yang}},\ }\href
  {https://doi.org/10.1007/JHEP09(2013)043} {\bibfield  {journal} {\bibinfo
  {journal} {JHEP}\ }\textbf {\bibinfo {volume} {09}},\ \bibinfo {pages}
  {043}},\ \Eprint {https://arxiv.org/abs/1301.4641} {arXiv:1301.4641 [hep-ph]}
  \BibitemShut {NoStop}%
\bibitem [{\citenamefont {Maru}\ and\ \citenamefont
  {Okada}(2013)}]{Maru:2013bja}%
  \BibitemOpen
  \bibfield  {author} {\bibinfo {author} {\bibfnamefont {N.}~\bibnamefont
  {Maru}}\ and\ \bibinfo {author} {\bibfnamefont {N.}~\bibnamefont {Okada}},\
  }\href {https://doi.org/10.1103/PhysRevD.88.037701} {\bibfield  {journal}
  {\bibinfo  {journal} {Phys. Rev. D}\ }\textbf {\bibinfo {volume} {88}},\
  \bibinfo {pages} {037701} (\bibinfo {year} {2013})},\ \Eprint
  {https://arxiv.org/abs/1307.0291} {arXiv:1307.0291 [hep-ph]} \BibitemShut
  {NoStop}%
\bibitem [{\citenamefont {Yue}\ \emph {et~al.}(2013)\citenamefont {Yue},
  \citenamefont {Shi},\ and\ \citenamefont {Hua}}]{Yue:2013qba}%
  \BibitemOpen
  \bibfield  {author} {\bibinfo {author} {\bibfnamefont {C.-X.}\ \bibnamefont
  {Yue}}, \bibinfo {author} {\bibfnamefont {Q.-Y.}\ \bibnamefont {Shi}},\ and\
  \bibinfo {author} {\bibfnamefont {T.}~\bibnamefont {Hua}},\ }\href
  {https://doi.org/10.1016/j.nuclphysb.2013.09.004} {\bibfield  {journal}
  {\bibinfo  {journal} {Nucl. Phys. B}\ }\textbf {\bibinfo {volume} {876}},\
  \bibinfo {pages} {747} (\bibinfo {year} {2013})},\ \Eprint
  {https://arxiv.org/abs/1307.5572} {arXiv:1307.5572 [hep-ph]} \BibitemShut
  {NoStop}%
\bibitem [{\citenamefont {Belanger}\ \emph {et~al.}(2014)\citenamefont
  {Belanger}, \citenamefont {Bizouard},\ and\ \citenamefont
  {Chalons}}]{Belanger:2014roa}%
  \BibitemOpen
  \bibfield  {author} {\bibinfo {author} {\bibfnamefont {G.}~\bibnamefont
  {Belanger}}, \bibinfo {author} {\bibfnamefont {V.}~\bibnamefont {Bizouard}},\
  and\ \bibinfo {author} {\bibfnamefont {G.}~\bibnamefont {Chalons}},\ }\href
  {https://doi.org/10.1103/PhysRevD.89.095023} {\bibfield  {journal} {\bibinfo
  {journal} {Phys. Rev. D}\ }\textbf {\bibinfo {volume} {89}},\ \bibinfo
  {pages} {095023} (\bibinfo {year} {2014})},\ \Eprint
  {https://arxiv.org/abs/1402.3522} {arXiv:1402.3522 [hep-ph]} \BibitemShut
  {NoStop}%
\bibitem [{\citenamefont {Fontes}\ \emph {et~al.}(2014)\citenamefont {Fontes},
  \citenamefont {Rom\~ao},\ and\ \citenamefont {Silva}}]{Fontes:2014xva}%
  \BibitemOpen
  \bibfield  {author} {\bibinfo {author} {\bibfnamefont {D.}~\bibnamefont
  {Fontes}}, \bibinfo {author} {\bibfnamefont {J.}~\bibnamefont {Rom\~ao}},\
  and\ \bibinfo {author} {\bibfnamefont {J.~a.~P.}\ \bibnamefont {Silva}},\
  }\href {https://doi.org/10.1007/JHEP12(2014)043} {\bibfield  {journal}
  {\bibinfo  {journal} {JHEP}\ }\textbf {\bibinfo {volume} {12}},\ \bibinfo
  {pages} {043}},\ \Eprint {https://arxiv.org/abs/1408.2534} {arXiv:1408.2534
  [hep-ph]} \BibitemShut {NoStop}%
\bibitem [{\citenamefont {Hammad}\ \emph {et~al.}(2015)\citenamefont {Hammad},
  \citenamefont {Khalil},\ and\ \citenamefont {Moretti}}]{Hammad:2015eca}%
  \BibitemOpen
  \bibfield  {author} {\bibinfo {author} {\bibfnamefont {A.}~\bibnamefont
  {Hammad}}, \bibinfo {author} {\bibfnamefont {S.}~\bibnamefont {Khalil}},\
  and\ \bibinfo {author} {\bibfnamefont {S.}~\bibnamefont {Moretti}},\ }\href
  {https://doi.org/10.1103/PhysRevD.92.095008} {\bibfield  {journal} {\bibinfo
  {journal} {Phys. Rev. D}\ }\textbf {\bibinfo {volume} {92}},\ \bibinfo
  {pages} {095008} (\bibinfo {year} {2015})},\ \Eprint
  {https://arxiv.org/abs/1503.05408} {arXiv:1503.05408 [hep-ph]} \BibitemShut
  {NoStop}%
\bibitem [{\citenamefont {Hung}\ \emph {et~al.}(2019)\citenamefont {Hung},
  \citenamefont {Hong}, \citenamefont {Phuong}, \citenamefont {Mai},\ and\
  \citenamefont {Hue}}]{Hung:2019jue}%
  \BibitemOpen
  \bibfield  {author} {\bibinfo {author} {\bibfnamefont {H.}~\bibnamefont
  {Hung}}, \bibinfo {author} {\bibfnamefont {T.}~\bibnamefont {Hong}}, \bibinfo
  {author} {\bibfnamefont {H.}~\bibnamefont {Phuong}}, \bibinfo {author}
  {\bibfnamefont {H.}~\bibnamefont {Mai}},\ and\ \bibinfo {author}
  {\bibfnamefont {L.}~\bibnamefont {Hue}},\ }\href
  {https://doi.org/10.1103/PhysRevD.100.075014} {\bibfield  {journal} {\bibinfo
   {journal} {Phys. Rev. D}\ }\textbf {\bibinfo {volume} {100}},\ \bibinfo
  {pages} {075014} (\bibinfo {year} {2019})},\ \Eprint
  {https://arxiv.org/abs/1907.06735} {arXiv:1907.06735 [hep-ph]} \BibitemShut
  {NoStop}%
\bibitem [{\citenamefont {Liu}\ \emph {et~al.}(2020{\natexlab{a}})\citenamefont
  {Liu}, \citenamefont {Zhang}, \citenamefont {Yang}, \citenamefont {Zhao},
  \citenamefont {Liu},\ and\ \citenamefont {Feng}}]{Liu:2020nsm}%
  \BibitemOpen
  \bibfield  {author} {\bibinfo {author} {\bibfnamefont {C.-X.}\ \bibnamefont
  {Liu}}, \bibinfo {author} {\bibfnamefont {H.-B.}\ \bibnamefont {Zhang}},
  \bibinfo {author} {\bibfnamefont {J.-L.}\ \bibnamefont {Yang}}, \bibinfo
  {author} {\bibfnamefont {S.-M.}\ \bibnamefont {Zhao}}, \bibinfo {author}
  {\bibfnamefont {Y.-B.}\ \bibnamefont {Liu}},\ and\ \bibinfo {author}
  {\bibfnamefont {T.-F.}\ \bibnamefont {Feng}},\ }\href
  {https://doi.org/10.1007/JHEP04(2020)002} {\bibfield  {journal} {\bibinfo
  {journal} {JHEP}\ }\textbf {\bibinfo {volume} {04}},\ \bibinfo {pages}
  {002}},\ \Eprint {https://arxiv.org/abs/2002.04370} {arXiv:2002.04370
  [hep-ph]} \BibitemShut {NoStop}%
\bibitem [{\citenamefont {He}(2020)}]{He:2020suf}%
  \BibitemOpen
  \bibfield  {author} {\bibinfo {author} {\bibfnamefont {S.-P.}\ \bibnamefont
  {He}},\ }\href {https://doi.org/10.1103/PhysRevD.102.075035} {\bibfield
  {journal} {\bibinfo  {journal} {Phys. Rev. D}\ }\textbf {\bibinfo {volume}
  {102}},\ \bibinfo {pages} {075035} (\bibinfo {year} {2020})},\ \Eprint
  {https://arxiv.org/abs/2004.12155} {arXiv:2004.12155 [hep-ph]} \BibitemShut
  {NoStop}%
\bibitem [{\citenamefont {Bhupal~Dev}\ \emph {et~al.}(2013)\citenamefont
  {Bhupal~Dev}, \citenamefont {Ghosh}, \citenamefont {Okada},\ and\
  \citenamefont {Saha}}]{Dev:2013ff}%
  \BibitemOpen
  \bibfield  {author} {\bibinfo {author} {\bibfnamefont {P.}~\bibnamefont
  {Bhupal~Dev}}, \bibinfo {author} {\bibfnamefont {D.~K.}\ \bibnamefont
  {Ghosh}}, \bibinfo {author} {\bibfnamefont {N.}~\bibnamefont {Okada}},\ and\
  \bibinfo {author} {\bibfnamefont {I.}~\bibnamefont {Saha}},\ }\href
  {https://doi.org/10.1007/JHEP03(2013)150} {\bibfield  {journal} {\bibinfo
  {journal} {JHEP}\ }\textbf {\bibinfo {volume} {03}},\ \bibinfo {pages}
  {150}},\ \bibinfo {note} {[Erratum: JHEP 05, 049 (2013)]},\ \Eprint
  {https://arxiv.org/abs/1301.3453} {arXiv:1301.3453 [hep-ph]} \BibitemShut
  {NoStop}%
\bibitem [{\citenamefont {Cao}\ \emph {et~al.}(2019)\citenamefont {Cao},
  \citenamefont {Xu}, \citenamefont {Yan},\ and\ \citenamefont
  {Zhu}}]{Cao:2018cms}%
  \BibitemOpen
  \bibfield  {author} {\bibinfo {author} {\bibfnamefont {Q.-H.}\ \bibnamefont
  {Cao}}, \bibinfo {author} {\bibfnamefont {L.-X.}\ \bibnamefont {Xu}},
  \bibinfo {author} {\bibfnamefont {B.}~\bibnamefont {Yan}},\ and\ \bibinfo
  {author} {\bibfnamefont {S.-H.}\ \bibnamefont {Zhu}},\ }\href
  {https://doi.org/10.1016/j.physletb.2018.12.040} {\bibfield  {journal}
  {\bibinfo  {journal} {Phys. Lett. B}\ }\textbf {\bibinfo {volume} {789}},\
  \bibinfo {pages} {233} (\bibinfo {year} {2019})},\ \Eprint
  {https://arxiv.org/abs/1810.07661} {arXiv:1810.07661 [hep-ph]} \BibitemShut
  {NoStop}%
\bibitem [{\citenamefont {Bizot}\ and\ \citenamefont
  {Frigerio}(2016)}]{Bizot:2015zaa}%
  \BibitemOpen
  \bibfield  {author} {\bibinfo {author} {\bibfnamefont {N.}~\bibnamefont
  {Bizot}}\ and\ \bibinfo {author} {\bibfnamefont {M.}~\bibnamefont
  {Frigerio}},\ }\href {https://doi.org/10.1007/JHEP01(2016)036} {\bibfield
  {journal} {\bibinfo  {journal} {JHEP}\ }\textbf {\bibinfo {volume} {01}},\
  \bibinfo {pages} {036}},\ \Eprint {https://arxiv.org/abs/1508.01645}
  {arXiv:1508.01645 [hep-ph]} \BibitemShut {NoStop}%
\bibitem [{\citenamefont {No}\ and\ \citenamefont
  {Spannowsky}(2017)}]{No:2016ezr}%
  \BibitemOpen
  \bibfield  {author} {\bibinfo {author} {\bibfnamefont {J.~M.}\ \bibnamefont
  {No}}\ and\ \bibinfo {author} {\bibfnamefont {M.}~\bibnamefont
  {Spannowsky}},\ }\href {https://doi.org/10.1103/PhysRevD.95.075027}
  {\bibfield  {journal} {\bibinfo  {journal} {Phys. Rev. D}\ }\textbf {\bibinfo
  {volume} {95}},\ \bibinfo {pages} {075027} (\bibinfo {year} {2017})},\
  \Eprint {https://arxiv.org/abs/1612.06626} {arXiv:1612.06626 [hep-ph]}
  \BibitemShut {NoStop}%
\bibitem [{\citenamefont {Cao}\ \emph {et~al.}(2015)\citenamefont {Cao},
  \citenamefont {Wang},\ and\ \citenamefont {Zhang}}]{Cao:2015iua}%
  \BibitemOpen
  \bibfield  {author} {\bibinfo {author} {\bibfnamefont {Q.-H.}\ \bibnamefont
  {Cao}}, \bibinfo {author} {\bibfnamefont {H.-R.}\ \bibnamefont {Wang}},\ and\
  \bibinfo {author} {\bibfnamefont {Y.}~\bibnamefont {Zhang}},\ }\href
  {https://doi.org/10.1088/1674-1137/39/11/113102} {\bibfield  {journal}
  {\bibinfo  {journal} {Chin. Phys. C}\ }\textbf {\bibinfo {volume} {39}},\
  \bibinfo {pages} {113102} (\bibinfo {year} {2015})},\ \Eprint
  {https://arxiv.org/abs/1505.00654} {arXiv:1505.00654 [hep-ph]} \BibitemShut
  {NoStop}%
\bibitem [{\citenamefont {Ellis}\ \emph {et~al.}(1976)\citenamefont {Ellis},
  \citenamefont {Gaillard},\ and\ \citenamefont {Nanopoulos}}]{Ellis:1975ap}%
  \BibitemOpen
  \bibfield  {author} {\bibinfo {author} {\bibfnamefont {J.~R.}\ \bibnamefont
  {Ellis}}, \bibinfo {author} {\bibfnamefont {M.~K.}\ \bibnamefont
  {Gaillard}},\ and\ \bibinfo {author} {\bibfnamefont {D.~V.}\ \bibnamefont
  {Nanopoulos}},\ }\href {https://doi.org/10.1016/0550-3213(76)90382-5}
  {\bibfield  {journal} {\bibinfo  {journal} {Nucl. Phys.}\ }\textbf {\bibinfo
  {volume} {B106}},\ \bibinfo {pages} {292} (\bibinfo {year}
  {1976})}\BibitemShut {NoStop}%
%%CITATION = NUPHA,B106,292;%%
\bibitem [{\citenamefont {Shifman}\ \emph {et~al.}(1979)\citenamefont
  {Shifman}, \citenamefont {Vainshtein}, \citenamefont {Voloshin},\ and\
  \citenamefont {Zakharov}}]{shifman1979low}%
  \BibitemOpen
  \bibfield  {author} {\bibinfo {author} {\bibfnamefont {M.~A.}\ \bibnamefont
  {Shifman}}, \bibinfo {author} {\bibfnamefont {A.}~\bibnamefont {Vainshtein}},
  \bibinfo {author} {\bibfnamefont {M.}~\bibnamefont {Voloshin}},\ and\
  \bibinfo {author} {\bibfnamefont {V.~I.}\ \bibnamefont {Zakharov}},\
  }\href@noop {} {\bibfield  {journal} {\bibinfo  {journal} {Sov. J. Nucl.
  Phys.}\ }\textbf {\bibinfo {volume} {30}},\ \bibinfo {pages} {1368} (\bibinfo
  {year} {1979})}\BibitemShut {NoStop}%
\bibitem [{\citenamefont {Ross}\ and\ \citenamefont
  {Veltman}(1975)}]{Ross:1975fq}%
  \BibitemOpen
  \bibfield  {author} {\bibinfo {author} {\bibfnamefont {D.}~\bibnamefont
  {Ross}}\ and\ \bibinfo {author} {\bibfnamefont {M.}~\bibnamefont {Veltman}},\
  }\href {https://doi.org/10.1016/0550-3213(75)90485-X} {\bibfield  {journal}
  {\bibinfo  {journal} {Nucl. Phys. B}\ }\textbf {\bibinfo {volume} {95}},\
  \bibinfo {pages} {135} (\bibinfo {year} {1975})}\BibitemShut {NoStop}%
\bibitem [{\citenamefont {Veltman}(1977)}]{Veltman:1977kh}%
  \BibitemOpen
  \bibfield  {author} {\bibinfo {author} {\bibfnamefont {M.}~\bibnamefont
  {Veltman}},\ }\href {https://doi.org/10.1016/0550-3213(77)90342-X} {\bibfield
   {journal} {\bibinfo  {journal} {Nucl. Phys. B}\ }\textbf {\bibinfo {volume}
  {123}},\ \bibinfo {pages} {89} (\bibinfo {year} {1977})}\BibitemShut
  {NoStop}%
\bibitem [{\citenamefont {Zyla}\ \emph {et~al.}(2020)\citenamefont {Zyla} \emph
  {et~al.}}]{Zyla:2020zbs}%
  \BibitemOpen
  \bibfield  {author} {\bibinfo {author} {\bibfnamefont {P.}~\bibnamefont
  {Zyla}} \emph {et~al.} (\bibinfo {collaboration} {Particle Data Group}),\
  }\href {https://doi.org/10.1093/ptep/ptaa104} {\bibfield  {journal} {\bibinfo
   {journal} {PTEP}\ }\textbf {\bibinfo {volume} {2020}},\ \bibinfo {pages}
  {083C01} (\bibinfo {year} {2020})}\BibitemShut {NoStop}%
\bibitem [{\citenamefont {Braibant}(2003)}]{Braibant:2003px}%
  \BibitemOpen
  \bibfield  {author} {\bibinfo {author} {\bibfnamefont {S.}~\bibnamefont
  {Braibant}},\ }in\ \href@noop {} {\emph {\bibinfo {booktitle} {{38th
  Rencontres de Moriond on QCD and High-Energy Hadronic Interactions}}}}\
  (\bibinfo {year} {2003})\ \Eprint {https://arxiv.org/abs/hep-ex/0305058}
  {arXiv:hep-ex/0305058} \BibitemShut {NoStop}%
\bibitem [{\citenamefont {Heister}\ \emph {et~al.}(2002)\citenamefont {Heister}
  \emph {et~al.}}]{Heister:2002hp}%
  \BibitemOpen
  \bibfield  {author} {\bibinfo {author} {\bibfnamefont {A.}~\bibnamefont
  {Heister}} \emph {et~al.} (\bibinfo {collaboration} {ALEPH}),\ }\href
  {https://doi.org/10.1016/S0370-2693(02)01827-0} {\bibfield  {journal}
  {\bibinfo  {journal} {Phys. Lett. B}\ }\textbf {\bibinfo {volume} {537}},\
  \bibinfo {pages} {5} (\bibinfo {year} {2002})},\ \Eprint
  {https://arxiv.org/abs/hep-ex/0204036} {arXiv:hep-ex/0204036} \BibitemShut
  {NoStop}%
\bibitem [{\citenamefont {Abdallah}\ \emph {et~al.}(2003)\citenamefont
  {Abdallah} \emph {et~al.}}]{Abdallah:2003xe}%
  \BibitemOpen
  \bibfield  {author} {\bibinfo {author} {\bibfnamefont {J.}~\bibnamefont
  {Abdallah}} \emph {et~al.} (\bibinfo {collaboration} {DELPHI}),\ }\href
  {https://doi.org/10.1140/epjc/s2003-01355-5} {\bibfield  {journal} {\bibinfo
  {journal} {Eur. Phys. J. C}\ }\textbf {\bibinfo {volume} {31}},\ \bibinfo
  {pages} {421} (\bibinfo {year} {2003})},\ \Eprint
  {https://arxiv.org/abs/hep-ex/0311019} {arXiv:hep-ex/0311019} \BibitemShut
  {NoStop}%
\bibitem [{\citenamefont {Achard}\ \emph {et~al.}(2004)\citenamefont {Achard}
  \emph {et~al.}}]{Achard:2003ge}%
  \BibitemOpen
  \bibfield  {author} {\bibinfo {author} {\bibfnamefont {P.}~\bibnamefont
  {Achard}} \emph {et~al.} (\bibinfo {collaboration} {L3}),\ }\href
  {https://doi.org/10.1016/j.physletb.2003.10.010} {\bibfield  {journal}
  {\bibinfo  {journal} {Phys. Lett. B}\ }\textbf {\bibinfo {volume} {580}},\
  \bibinfo {pages} {37} (\bibinfo {year} {2004})},\ \Eprint
  {https://arxiv.org/abs/hep-ex/0310007} {arXiv:hep-ex/0310007} \BibitemShut
  {NoStop}%
\bibitem [{\citenamefont {Abbiendi}\ \emph {et~al.}(2002)\citenamefont
  {Abbiendi} \emph {et~al.}}]{Abbiendi:2002mp}%
  \BibitemOpen
  \bibfield  {author} {\bibinfo {author} {\bibfnamefont {G.}~\bibnamefont
  {Abbiendi}} \emph {et~al.} (\bibinfo {collaboration} {OPAL}),\ }\href
  {https://doi.org/10.1016/S0370-2693(02)02593-5} {\bibfield  {journal}
  {\bibinfo  {journal} {Phys. Lett. B}\ }\textbf {\bibinfo {volume} {545}},\
  \bibinfo {pages} {272} (\bibinfo {year} {2002})},\ \bibinfo {note} {[Erratum:
  Phys.Lett.B 548, 258--258 (2002)]},\ \Eprint
  {https://arxiv.org/abs/hep-ex/0209026} {arXiv:hep-ex/0209026} \BibitemShut
  {NoStop}%
\bibitem [{\citenamefont {Bolognesi}\ \emph {et~al.}(2012)\citenamefont
  {Bolognesi}, \citenamefont {Gao}, \citenamefont {Gritsan}, \citenamefont
  {Melnikov}, \citenamefont {Schulze}, \citenamefont {Tran},\ and\
  \citenamefont {Whitbeck}}]{Bolognesi:2012mm}%
  \BibitemOpen
  \bibfield  {author} {\bibinfo {author} {\bibfnamefont {S.}~\bibnamefont
  {Bolognesi}}, \bibinfo {author} {\bibfnamefont {Y.}~\bibnamefont {Gao}},
  \bibinfo {author} {\bibfnamefont {A.~V.}\ \bibnamefont {Gritsan}}, \bibinfo
  {author} {\bibfnamefont {K.}~\bibnamefont {Melnikov}}, \bibinfo {author}
  {\bibfnamefont {M.}~\bibnamefont {Schulze}}, \bibinfo {author} {\bibfnamefont
  {N.~V.}\ \bibnamefont {Tran}},\ and\ \bibinfo {author} {\bibfnamefont
  {A.}~\bibnamefont {Whitbeck}},\ }\href
  {https://doi.org/10.1103/PhysRevD.86.095031} {\bibfield  {journal} {\bibinfo
  {journal} {Phys. Rev. D}\ }\textbf {\bibinfo {volume} {86}},\ \bibinfo
  {pages} {095031} (\bibinfo {year} {2012})},\ \Eprint
  {https://arxiv.org/abs/1208.4018} {arXiv:1208.4018 [hep-ph]} \BibitemShut
  {NoStop}%
\bibitem [{\citenamefont {Stolarski}\ and\ \citenamefont
  {Vega-Morales}(2012)}]{Stolarski:2012ps}%
  \BibitemOpen
  \bibfield  {author} {\bibinfo {author} {\bibfnamefont {D.}~\bibnamefont
  {Stolarski}}\ and\ \bibinfo {author} {\bibfnamefont {R.}~\bibnamefont
  {Vega-Morales}},\ }\href {https://doi.org/10.1103/PhysRevD.86.117504}
  {\bibfield  {journal} {\bibinfo  {journal} {Phys. Rev. D}\ }\textbf {\bibinfo
  {volume} {86}},\ \bibinfo {pages} {117504} (\bibinfo {year} {2012})},\
  \Eprint {https://arxiv.org/abs/1208.4840} {arXiv:1208.4840 [hep-ph]}
  \BibitemShut {NoStop}%
\bibitem [{\citenamefont {Chen}\ and\ \citenamefont
  {Vega-Morales}(2014)}]{Chen:2013ejz}%
  \BibitemOpen
  \bibfield  {author} {\bibinfo {author} {\bibfnamefont {Y.}~\bibnamefont
  {Chen}}\ and\ \bibinfo {author} {\bibfnamefont {R.}~\bibnamefont
  {Vega-Morales}},\ }\href {https://doi.org/10.1007/JHEP04(2014)057} {\bibfield
   {journal} {\bibinfo  {journal} {JHEP}\ }\textbf {\bibinfo {volume} {04}},\
  \bibinfo {pages} {057}},\ \Eprint {https://arxiv.org/abs/1310.2893}
  {arXiv:1310.2893 [hep-ph]} \BibitemShut {NoStop}%
\bibitem [{\citenamefont {Chen}\ \emph
  {et~al.}(2014{\natexlab{a}})\citenamefont {Chen}, \citenamefont {Harnik},\
  and\ \citenamefont {Vega-Morales}}]{Chen:2014gka}%
  \BibitemOpen
  \bibfield  {author} {\bibinfo {author} {\bibfnamefont {Y.}~\bibnamefont
  {Chen}}, \bibinfo {author} {\bibfnamefont {R.}~\bibnamefont {Harnik}},\ and\
  \bibinfo {author} {\bibfnamefont {R.}~\bibnamefont {Vega-Morales}},\ }\href
  {https://doi.org/10.1103/PhysRevLett.113.191801} {\bibfield  {journal}
  {\bibinfo  {journal} {Phys. Rev. Lett.}\ }\textbf {\bibinfo {volume} {113}},\
  \bibinfo {pages} {191801} (\bibinfo {year} {2014}{\natexlab{a}})},\ \Eprint
  {https://arxiv.org/abs/1404.1336} {arXiv:1404.1336 [hep-ph]} \BibitemShut
  {NoStop}%
\bibitem [{\citenamefont {Falkowski}\ and\ \citenamefont
  {Vega-Morales}(2014)}]{Falkowski:2014ffa}%
  \BibitemOpen
  \bibfield  {author} {\bibinfo {author} {\bibfnamefont {A.}~\bibnamefont
  {Falkowski}}\ and\ \bibinfo {author} {\bibfnamefont {R.}~\bibnamefont
  {Vega-Morales}},\ }\href {https://doi.org/10.1007/JHEP12(2014)037} {\bibfield
   {journal} {\bibinfo  {journal} {JHEP}\ }\textbf {\bibinfo {volume} {12}},\
  \bibinfo {pages} {037}},\ \Eprint {https://arxiv.org/abs/1405.1095}
  {arXiv:1405.1095 [hep-ph]} \BibitemShut {NoStop}%
\bibitem [{\citenamefont {Chen}\ \emph
  {et~al.}(2015{\natexlab{a}})\citenamefont {Chen}, \citenamefont {Stolarski},\
  and\ \citenamefont {Vega-Morales}}]{Chen:2015rha}%
  \BibitemOpen
  \bibfield  {author} {\bibinfo {author} {\bibfnamefont {Y.}~\bibnamefont
  {Chen}}, \bibinfo {author} {\bibfnamefont {D.}~\bibnamefont {Stolarski}},\
  and\ \bibinfo {author} {\bibfnamefont {R.}~\bibnamefont {Vega-Morales}},\
  }\href {https://doi.org/10.1103/PhysRevD.92.053003} {\bibfield  {journal}
  {\bibinfo  {journal} {Phys. Rev.}\ }\textbf {\bibinfo {volume} {D92}},\
  \bibinfo {pages} {053003} (\bibinfo {year} {2015}{\natexlab{a}})},\ \Eprint
  {https://arxiv.org/abs/1505.01168} {arXiv:1505.01168 [hep-ph]} \BibitemShut
  {NoStop}%
%%CITATION = ARXIV:1505.01168;%%
\bibitem [{\citenamefont {Chen}\ \emph {et~al.}(2016)\citenamefont {Chen},
  \citenamefont {Lykken}, \citenamefont {Spiropulu}, \citenamefont
  {Stolarski},\ and\ \citenamefont {Vega-Morales}}]{Chen:2016ofc}%
  \BibitemOpen
  \bibfield  {author} {\bibinfo {author} {\bibfnamefont {Y.}~\bibnamefont
  {Chen}}, \bibinfo {author} {\bibfnamefont {J.}~\bibnamefont {Lykken}},
  \bibinfo {author} {\bibfnamefont {M.}~\bibnamefont {Spiropulu}}, \bibinfo
  {author} {\bibfnamefont {D.}~\bibnamefont {Stolarski}},\ and\ \bibinfo
  {author} {\bibfnamefont {R.}~\bibnamefont {Vega-Morales}},\ }\href
  {https://doi.org/10.1103/PhysRevLett.117.241801} {\bibfield  {journal}
  {\bibinfo  {journal} {Phys. Rev. Lett.}\ }\textbf {\bibinfo {volume} {117}},\
  \bibinfo {pages} {241801} (\bibinfo {year} {2016})},\ \Eprint
  {https://arxiv.org/abs/1608.02159} {arXiv:1608.02159 [hep-ph]} \BibitemShut
  {NoStop}%
%%CITATION = ARXIV:1608.02159;%%
\bibitem [{\citenamefont {Boselli}\ \emph {et~al.}(2018)\citenamefont
  {Boselli}, \citenamefont {Carloni~Calame}, \citenamefont {Montagna},
  \citenamefont {Nicrosini}, \citenamefont {Piccinini},\ and\ \citenamefont
  {Shivaji}}]{Boselli:2017pef}%
  \BibitemOpen
  \bibfield  {author} {\bibinfo {author} {\bibfnamefont {S.}~\bibnamefont
  {Boselli}}, \bibinfo {author} {\bibfnamefont {C.~M.}\ \bibnamefont
  {Carloni~Calame}}, \bibinfo {author} {\bibfnamefont {G.}~\bibnamefont
  {Montagna}}, \bibinfo {author} {\bibfnamefont {O.}~\bibnamefont {Nicrosini}},
  \bibinfo {author} {\bibfnamefont {F.}~\bibnamefont {Piccinini}},\ and\
  \bibinfo {author} {\bibfnamefont {A.}~\bibnamefont {Shivaji}},\ }\href
  {https://doi.org/10.1007/JHEP01(2018)096} {\bibfield  {journal} {\bibinfo
  {journal} {JHEP}\ }\textbf {\bibinfo {volume} {01}},\ \bibinfo {pages}
  {096}},\ \Eprint {https://arxiv.org/abs/1703.06667} {arXiv:1703.06667
  [hep-ph]} \BibitemShut {NoStop}%
\bibitem [{\citenamefont {Anderson}\ \emph {et~al.}(2014)\citenamefont
  {Anderson} \emph {et~al.}}]{Anderson:2013afp}%
  \BibitemOpen
  \bibfield  {author} {\bibinfo {author} {\bibfnamefont {I.}~\bibnamefont
  {Anderson}} \emph {et~al.},\ }\href
  {https://doi.org/10.1103/PhysRevD.89.035007} {\bibfield  {journal} {\bibinfo
  {journal} {Phys. Rev. D}\ }\textbf {\bibinfo {volume} {89}},\ \bibinfo
  {pages} {035007} (\bibinfo {year} {2014})},\ \Eprint
  {https://arxiv.org/abs/1309.4819} {arXiv:1309.4819 [hep-ph]} \BibitemShut
  {NoStop}%
\bibitem [{\citenamefont {Gainer}\ \emph {et~al.}(2015)\citenamefont {Gainer},
  \citenamefont {Lykken}, \citenamefont {Matchev}, \citenamefont {Mrenna},\
  and\ \citenamefont {Park}}]{Gainer:2014hha}%
  \BibitemOpen
  \bibfield  {author} {\bibinfo {author} {\bibfnamefont {J.~S.}\ \bibnamefont
  {Gainer}}, \bibinfo {author} {\bibfnamefont {J.}~\bibnamefont {Lykken}},
  \bibinfo {author} {\bibfnamefont {K.~T.}\ \bibnamefont {Matchev}}, \bibinfo
  {author} {\bibfnamefont {S.}~\bibnamefont {Mrenna}},\ and\ \bibinfo {author}
  {\bibfnamefont {M.}~\bibnamefont {Park}},\ }\href
  {https://doi.org/10.1103/PhysRevD.91.035011} {\bibfield  {journal} {\bibinfo
  {journal} {Phys. Rev.}\ }\textbf {\bibinfo {volume} {D91}},\ \bibinfo {pages}
  {035011} (\bibinfo {year} {2015})},\ \Eprint
  {https://arxiv.org/abs/1403.4951} {arXiv:1403.4951 [hep-ph]} \BibitemShut
  {NoStop}%
%%CITATION = ARXIV:1403.4951;%%
\bibitem [{\citenamefont {Liu}\ \emph {et~al.}(2020{\natexlab{b}})\citenamefont
  {Liu}, \citenamefont {Low},\ and\ \citenamefont
  {Vega-Morales}}]{Liu:2019rce}%
  \BibitemOpen
  \bibfield  {author} {\bibinfo {author} {\bibfnamefont {D.}~\bibnamefont
  {Liu}}, \bibinfo {author} {\bibfnamefont {I.}~\bibnamefont {Low}},\ and\
  \bibinfo {author} {\bibfnamefont {R.}~\bibnamefont {Vega-Morales}},\ }\href
  {https://doi.org/10.1140/epjc/s10052-020-8244-8} {\bibfield  {journal}
  {\bibinfo  {journal} {Eur. Phys. J. C}\ }\textbf {\bibinfo {volume} {80}},\
  \bibinfo {pages} {829} (\bibinfo {year} {2020}{\natexlab{b}})},\ \Eprint
  {https://arxiv.org/abs/1904.00026} {arXiv:1904.00026 [hep-ph]} \BibitemShut
  {NoStop}%
\bibitem [{\citenamefont {Sirunyan}\ \emph {et~al.}(2017)\citenamefont
  {Sirunyan} \emph {et~al.}}]{Sirunyan:2017tqd}%
  \BibitemOpen
  \bibfield  {author} {\bibinfo {author} {\bibfnamefont {A.~M.}\ \bibnamefont
  {Sirunyan}} \emph {et~al.} (\bibinfo {collaboration} {CMS}),\ }\href
  {https://doi.org/10.1016/j.physletb.2017.10.021} {\bibfield  {journal}
  {\bibinfo  {journal} {Phys. Lett. B}\ }\textbf {\bibinfo {volume} {775}},\
  \bibinfo {pages} {1} (\bibinfo {year} {2017})},\ \Eprint
  {https://arxiv.org/abs/1707.00541} {arXiv:1707.00541 [hep-ex]} \BibitemShut
  {NoStop}%
\bibitem [{\citenamefont {Chen}\ \emph
  {et~al.}(2015{\natexlab{b}})\citenamefont {Chen}, \citenamefont {Harnik},\
  and\ \citenamefont {Vega-Morales}}]{Chen:2015iha}%
  \BibitemOpen
  \bibfield  {author} {\bibinfo {author} {\bibfnamefont {Y.}~\bibnamefont
  {Chen}}, \bibinfo {author} {\bibfnamefont {R.}~\bibnamefont {Harnik}},\ and\
  \bibinfo {author} {\bibfnamefont {R.}~\bibnamefont {Vega-Morales}},\ }\href
  {https://doi.org/10.1007/JHEP09(2015)185} {\bibfield  {journal} {\bibinfo
  {journal} {JHEP}\ }\textbf {\bibinfo {volume} {09}},\ \bibinfo {pages}
  {185}},\ \Eprint {https://arxiv.org/abs/1503.05855} {arXiv:1503.05855
  [hep-ph]} \BibitemShut {NoStop}%
%%CITATION = ARXIV:1503.05855;%%
\bibitem [{\citenamefont {Martin}(1997)}]{Martin:1997ns}%
  \BibitemOpen
  \bibfield  {author} {\bibinfo {author} {\bibfnamefont {S.~P.}\ \bibnamefont
  {Martin}},\ }\href {https://doi.org/10.1142/9789812839657_0001,
  10.1142/9789814307505_0001} {\ ,\ \bibinfo {pages} {1} (\bibinfo {year}
  {1997})},\ \bibinfo {note} {[Adv. Ser. Direct. High Energy
  Phys.18,1(1998)]},\ \Eprint {https://arxiv.org/abs/hep-ph/9709356}
  {arXiv:hep-ph/9709356 [hep-ph]} \BibitemShut {NoStop}%
%%CITATION = HEP-PH/9709356;%%
\bibitem [{\citenamefont {Fan}\ and\ \citenamefont
  {Reece}(2014)}]{Fan:2014txa}%
  \BibitemOpen
  \bibfield  {author} {\bibinfo {author} {\bibfnamefont {J.}~\bibnamefont
  {Fan}}\ and\ \bibinfo {author} {\bibfnamefont {M.}~\bibnamefont {Reece}},\
  }\href {https://doi.org/10.1007/JHEP06(2014)031} {\bibfield  {journal}
  {\bibinfo  {journal} {JHEP}\ }\textbf {\bibinfo {volume} {06}},\ \bibinfo
  {pages} {031}},\ \Eprint {https://arxiv.org/abs/1401.7671} {arXiv:1401.7671
  [hep-ph]} \BibitemShut {NoStop}%
%%CITATION = ARXIV:1401.7671;%%
\bibitem [{\citenamefont {Burdman}\ \emph {et~al.}(2007)\citenamefont
  {Burdman}, \citenamefont {Chacko}, \citenamefont {Goh},\ and\ \citenamefont
  {Harnik}}]{Burdman:2006tz}%
  \BibitemOpen
  \bibfield  {author} {\bibinfo {author} {\bibfnamefont {G.}~\bibnamefont
  {Burdman}}, \bibinfo {author} {\bibfnamefont {Z.}~\bibnamefont {Chacko}},
  \bibinfo {author} {\bibfnamefont {H.-S.}\ \bibnamefont {Goh}},\ and\ \bibinfo
  {author} {\bibfnamefont {R.}~\bibnamefont {Harnik}},\ }\href
  {https://doi.org/10.1088/1126-6708/2007/02/009} {\bibfield  {journal}
  {\bibinfo  {journal} {JHEP}\ }\textbf {\bibinfo {volume} {02}},\ \bibinfo
  {pages} {009}},\ \Eprint {https://arxiv.org/abs/hep-ph/0609152}
  {arXiv:hep-ph/0609152 [hep-ph]} \BibitemShut {NoStop}%
%%CITATION = HEP-PH/0609152;%%
\bibitem [{\citenamefont {Andersen}\ \emph {et~al.}(2013)\citenamefont
  {Andersen} \emph {et~al.}}]{Heinemeyer:2013tqa}%
  \BibitemOpen
  \bibfield  {author} {\bibinfo {author} {\bibfnamefont {J.~R.}\ \bibnamefont
  {Andersen}} \emph {et~al.} (\bibinfo {collaboration} {LHC Higgs Cross Section
  Working Group})\ }\href {https://doi.org/10.5170/CERN-2013-004}
  {10.5170/CERN-2013-004} (\bibinfo {year} {2013}),\ \Eprint
  {https://arxiv.org/abs/1307.1347} {arXiv:1307.1347 [hep-ph]} \BibitemShut
  {NoStop}%
\bibitem [{\citenamefont {Cepeda}\ \emph {et~al.}(2019)\citenamefont {Cepeda}
  \emph {et~al.}}]{HLLHC}%
  \BibitemOpen
  \bibfield  {author} {\bibinfo {author} {\bibfnamefont {M.}~\bibnamefont
  {Cepeda}} \emph {et~al.},\ }\href
  {https://doi.org/10.23731/CYRM-2019-007.221} {\bibfield  {journal} {\bibinfo
  {journal} {CERN Yellow Rep. Monogr.}\ }\textbf {\bibinfo {volume} {7}},\
  \bibinfo {pages} {221} (\bibinfo {year} {2019})},\ \Eprint
  {https://arxiv.org/abs/1902.00134} {arXiv:1902.00134 [hep-ph]} \BibitemShut
  {NoStop}%
\bibitem [{\citenamefont {Chen}\ \emph
  {et~al.}(2015{\natexlab{c}})\citenamefont {Chen}, \citenamefont {Di~Marco},
  \citenamefont {Lykken}, \citenamefont {Spiropulu}, \citenamefont
  {Vega-Morales},\ and\ \citenamefont {Xie}}]{Chen:2014pia}%
  \BibitemOpen
  \bibfield  {author} {\bibinfo {author} {\bibfnamefont {Y.}~\bibnamefont
  {Chen}}, \bibinfo {author} {\bibfnamefont {E.}~\bibnamefont {Di~Marco}},
  \bibinfo {author} {\bibfnamefont {J.}~\bibnamefont {Lykken}}, \bibinfo
  {author} {\bibfnamefont {M.}~\bibnamefont {Spiropulu}}, \bibinfo {author}
  {\bibfnamefont {R.}~\bibnamefont {Vega-Morales}},\ and\ \bibinfo {author}
  {\bibfnamefont {S.}~\bibnamefont {Xie}},\ }\href
  {https://doi.org/10.1007/JHEP01(2015)125} {\bibfield  {journal} {\bibinfo
  {journal} {JHEP}\ }\textbf {\bibinfo {volume} {01}},\ \bibinfo {pages}
  {125}},\ \Eprint {https://arxiv.org/abs/1401.2077} {arXiv:1401.2077 [hep-ex]}
  \BibitemShut {NoStop}%
\bibitem [{\citenamefont {Bredenstein}\ \emph {et~al.}(2006)\citenamefont
  {Bredenstein}, \citenamefont {Denner}, \citenamefont {Dittmaier},\ and\
  \citenamefont {Weber}}]{Bredenstein:2006rh}%
  \BibitemOpen
  \bibfield  {author} {\bibinfo {author} {\bibfnamefont {A.}~\bibnamefont
  {Bredenstein}}, \bibinfo {author} {\bibfnamefont {A.}~\bibnamefont {Denner}},
  \bibinfo {author} {\bibfnamefont {S.}~\bibnamefont {Dittmaier}},\ and\
  \bibinfo {author} {\bibfnamefont {M.~M.}\ \bibnamefont {Weber}},\ }\href
  {https://doi.org/10.1103/PhysRevD.74.013004} {\bibfield  {journal} {\bibinfo
  {journal} {Phys. Rev. D}\ }\textbf {\bibinfo {volume} {74}},\ \bibinfo
  {pages} {013004} (\bibinfo {year} {2006})},\ \Eprint
  {https://arxiv.org/abs/hep-ph/0604011} {arXiv:hep-ph/0604011} \BibitemShut
  {NoStop}%
\bibitem [{\citenamefont {Bredenstein}\ \emph {et~al.}(2007)\citenamefont
  {Bredenstein}, \citenamefont {Denner}, \citenamefont {Dittmaier},\ and\
  \citenamefont {Weber}}]{Bredenstein:2006ha}%
  \BibitemOpen
  \bibfield  {author} {\bibinfo {author} {\bibfnamefont {A.}~\bibnamefont
  {Bredenstein}}, \bibinfo {author} {\bibfnamefont {A.}~\bibnamefont {Denner}},
  \bibinfo {author} {\bibfnamefont {S.}~\bibnamefont {Dittmaier}},\ and\
  \bibinfo {author} {\bibfnamefont {M.~M.}\ \bibnamefont {Weber}},\ }\href
  {https://doi.org/10.1088/1126-6708/2007/02/080} {\bibfield  {journal}
  {\bibinfo  {journal} {JHEP}\ }\textbf {\bibinfo {volume} {02}},\ \bibinfo
  {pages} {080}},\ \Eprint {https://arxiv.org/abs/hep-ph/0611234}
  {arXiv:hep-ph/0611234} \BibitemShut {NoStop}%
\bibitem [{\citenamefont {Low}\ \emph {et~al.}(2012)\citenamefont {Low},
  \citenamefont {Lykken},\ and\ \citenamefont {Shaughnessy}}]{Low:2012rj}%
  \BibitemOpen
  \bibfield  {author} {\bibinfo {author} {\bibfnamefont {I.}~\bibnamefont
  {Low}}, \bibinfo {author} {\bibfnamefont {J.}~\bibnamefont {Lykken}},\ and\
  \bibinfo {author} {\bibfnamefont {G.}~\bibnamefont {Shaughnessy}},\ }\href
  {https://doi.org/10.1103/PhysRevD.86.093012} {\bibfield  {journal} {\bibinfo
  {journal} {Phys. Rev.}\ }\textbf {\bibinfo {volume} {D86}},\ \bibinfo {pages}
  {093012} (\bibinfo {year} {2012})},\ \Eprint
  {https://arxiv.org/abs/1207.1093} {arXiv:1207.1093 [hep-ph]} \BibitemShut
  {NoStop}%
%%CITATION = ARXIV:1207.1093;%%
\bibitem [{\citenamefont {Brod}\ \emph {et~al.}(2013)\citenamefont {Brod},
  \citenamefont {Haisch},\ and\ \citenamefont {Zupan}}]{Brod:2013cka}%
  \BibitemOpen
  \bibfield  {author} {\bibinfo {author} {\bibfnamefont {J.}~\bibnamefont
  {Brod}}, \bibinfo {author} {\bibfnamefont {U.}~\bibnamefont {Haisch}},\ and\
  \bibinfo {author} {\bibfnamefont {J.}~\bibnamefont {Zupan}},\ }\href
  {https://doi.org/10.1007/JHEP11(2013)180} {\bibfield  {journal} {\bibinfo
  {journal} {JHEP}\ }\textbf {\bibinfo {volume} {11}},\ \bibinfo {pages}
  {180}},\ \Eprint {https://arxiv.org/abs/1310.1385} {arXiv:1310.1385 [hep-ph]}
  \BibitemShut {NoStop}%
\bibitem [{\citenamefont {Farina}\ \emph {et~al.}(2015)\citenamefont {Farina},
  \citenamefont {Grossman},\ and\ \citenamefont {Robinson}}]{Farina:2015dua}%
  \BibitemOpen
  \bibfield  {author} {\bibinfo {author} {\bibfnamefont {M.}~\bibnamefont
  {Farina}}, \bibinfo {author} {\bibfnamefont {Y.}~\bibnamefont {Grossman}},\
  and\ \bibinfo {author} {\bibfnamefont {D.~J.}\ \bibnamefont {Robinson}},\
  }\href {https://doi.org/10.1103/PhysRevD.92.073007} {\bibfield  {journal}
  {\bibinfo  {journal} {Phys. Rev. D}\ }\textbf {\bibinfo {volume} {92}},\
  \bibinfo {pages} {073007} (\bibinfo {year} {2015})},\ \Eprint
  {https://arxiv.org/abs/1503.06470} {arXiv:1503.06470 [hep-ph]} \BibitemShut
  {NoStop}%
\bibitem [{\citenamefont {Chen}\ \emph {et~al.}(2017)\citenamefont {Chen},
  \citenamefont {Li},\ and\ \citenamefont {Wan}}]{Chen:2017plj}%
  \BibitemOpen
  \bibfield  {author} {\bibinfo {author} {\bibfnamefont {X.}~\bibnamefont
  {Chen}}, \bibinfo {author} {\bibfnamefont {G.}~\bibnamefont {Li}},\ and\
  \bibinfo {author} {\bibfnamefont {X.}~\bibnamefont {Wan}},\ }\href
  {https://doi.org/10.1103/PhysRevD.96.055023} {\bibfield  {journal} {\bibinfo
  {journal} {Phys. Rev. D}\ }\textbf {\bibinfo {volume} {96}},\ \bibinfo
  {pages} {055023} (\bibinfo {year} {2017})},\ \Eprint
  {https://arxiv.org/abs/1705.01254} {arXiv:1705.01254 [hep-ph]} \BibitemShut
  {NoStop}%
\bibitem [{\citenamefont {Chen}\ \emph
  {et~al.}(2014{\natexlab{b}})\citenamefont {Chen}, \citenamefont {Falkowski},
  \citenamefont {Low},\ and\ \citenamefont {Vega-Morales}}]{Chen:2014ona}%
  \BibitemOpen
  \bibfield  {author} {\bibinfo {author} {\bibfnamefont {Y.}~\bibnamefont
  {Chen}}, \bibinfo {author} {\bibfnamefont {A.}~\bibnamefont {Falkowski}},
  \bibinfo {author} {\bibfnamefont {I.}~\bibnamefont {Low}},\ and\ \bibinfo
  {author} {\bibfnamefont {R.}~\bibnamefont {Vega-Morales}},\ }\href
  {https://doi.org/10.1103/PhysRevD.90.113006} {\bibfield  {journal} {\bibinfo
  {journal} {Phys. Rev. D}\ }\textbf {\bibinfo {volume} {90}},\ \bibinfo
  {pages} {113006} (\bibinfo {year} {2014}{\natexlab{b}})},\ \Eprint
  {https://arxiv.org/abs/1405.6723} {arXiv:1405.6723 [hep-ph]} \BibitemShut
  {NoStop}%
\bibitem [{\citenamefont {Gonzalez-Alonso}\ \emph {et~al.}(2015)\citenamefont
  {Gonzalez-Alonso}, \citenamefont {Greljo}, \citenamefont {Isidori},\ and\
  \citenamefont {Marzocca}}]{Gonzalez-Alonso:2014eva}%
  \BibitemOpen
  \bibfield  {author} {\bibinfo {author} {\bibfnamefont {M.}~\bibnamefont
  {Gonzalez-Alonso}}, \bibinfo {author} {\bibfnamefont {A.}~\bibnamefont
  {Greljo}}, \bibinfo {author} {\bibfnamefont {G.}~\bibnamefont {Isidori}},\
  and\ \bibinfo {author} {\bibfnamefont {D.}~\bibnamefont {Marzocca}},\ }\href
  {https://doi.org/10.1140/epjc/s10052-015-3345-5} {\bibfield  {journal}
  {\bibinfo  {journal} {Eur. Phys. J.}\ }\textbf {\bibinfo {volume} {C75}},\
  \bibinfo {pages} {128} (\bibinfo {year} {2015})},\ \Eprint
  {https://arxiv.org/abs/1412.6038} {arXiv:1412.6038 [hep-ph]} \BibitemShut
  {NoStop}%
%%CITATION = ARXIV:1412.6038;%%
\bibitem [{\citenamefont {Alwall}\ \emph {et~al.}(2014)\citenamefont {Alwall},
  \citenamefont {Frederix}, \citenamefont {Frixione}, \citenamefont {Hirschi},
  \citenamefont {Maltoni}, \citenamefont {Mattelaer}, \citenamefont {Shao},
  \citenamefont {Stelzer}, \citenamefont {Torrielli},\ and\ \citenamefont
  {Zaro}}]{Alwall:2014hca}%
  \BibitemOpen
  \bibfield  {author} {\bibinfo {author} {\bibfnamefont {J.}~\bibnamefont
  {Alwall}}, \bibinfo {author} {\bibfnamefont {R.}~\bibnamefont {Frederix}},
  \bibinfo {author} {\bibfnamefont {S.}~\bibnamefont {Frixione}}, \bibinfo
  {author} {\bibfnamefont {V.}~\bibnamefont {Hirschi}}, \bibinfo {author}
  {\bibfnamefont {F.}~\bibnamefont {Maltoni}}, \bibinfo {author} {\bibfnamefont
  {O.}~\bibnamefont {Mattelaer}}, \bibinfo {author} {\bibfnamefont {H.~S.}\
  \bibnamefont {Shao}}, \bibinfo {author} {\bibfnamefont {T.}~\bibnamefont
  {Stelzer}}, \bibinfo {author} {\bibfnamefont {P.}~\bibnamefont {Torrielli}},\
  and\ \bibinfo {author} {\bibfnamefont {M.}~\bibnamefont {Zaro}},\ }\href
  {https://doi.org/10.1007/JHEP07(2014)079} {\bibfield  {journal} {\bibinfo
  {journal} {JHEP}\ }\textbf {\bibinfo {volume} {07}},\ \bibinfo {pages}
  {079}},\ \Eprint {https://arxiv.org/abs/1405.0301} {arXiv:1405.0301 [hep-ph]}
  \BibitemShut {NoStop}%
\bibitem [{\citenamefont {Dittmaier}\ \emph {et~al.}(2011)\citenamefont
  {Dittmaier} \emph {et~al.}}]{Dittmaier:2011ti}%
  \BibitemOpen
  \bibfield  {author} {\bibinfo {author} {\bibfnamefont {S.}~\bibnamefont
  {Dittmaier}} \emph {et~al.} (\bibinfo {collaboration} {LHC Higgs Cross
  Section Working Group})\ }\href {https://doi.org/10.5170/CERN-2011-002}
  {10.5170/CERN-2011-002} (\bibinfo {year} {2011}),\ \Eprint
  {https://arxiv.org/abs/1101.0593} {arXiv:1101.0593 [hep-ph]} \BibitemShut
  {NoStop}%
\bibitem [{\citenamefont {Sirunyan}\ \emph {et~al.}(2021)\citenamefont
  {Sirunyan} \emph {et~al.}}]{Sirunyan:2021rug}%
  \BibitemOpen
  \bibfield  {author} {\bibinfo {author} {\bibfnamefont {A.~M.}\ \bibnamefont
  {Sirunyan}} \emph {et~al.} (\bibinfo {collaboration} {CMS}),\ }\href@noop {}
  {\  (\bibinfo {year} {2021})},\ \Eprint {https://arxiv.org/abs/2103.04956}
  {arXiv:2103.04956 [hep-ex]} \BibitemShut {NoStop}%
\bibitem [{\citenamefont {Gao}\ \emph {et~al.}(2010)\citenamefont {Gao},
  \citenamefont {Gritsan}, \citenamefont {Guo}, \citenamefont {Melnikov},
  \citenamefont {Schulze},\ and\ \citenamefont {Tran}}]{Gao:2010qx}%
  \BibitemOpen
  \bibfield  {author} {\bibinfo {author} {\bibfnamefont {Y.}~\bibnamefont
  {Gao}}, \bibinfo {author} {\bibfnamefont {A.~V.}\ \bibnamefont {Gritsan}},
  \bibinfo {author} {\bibfnamefont {Z.}~\bibnamefont {Guo}}, \bibinfo {author}
  {\bibfnamefont {K.}~\bibnamefont {Melnikov}}, \bibinfo {author}
  {\bibfnamefont {M.}~\bibnamefont {Schulze}},\ and\ \bibinfo {author}
  {\bibfnamefont {N.~V.}\ \bibnamefont {Tran}},\ }\href
  {https://doi.org/10.1103/PhysRevD.81.075022} {\bibfield  {journal} {\bibinfo
  {journal} {Phys. Rev. D}\ }\textbf {\bibinfo {volume} {81}},\ \bibinfo
  {pages} {075022} (\bibinfo {year} {2010})},\ \Eprint
  {https://arxiv.org/abs/1001.3396} {arXiv:1001.3396 [hep-ph]} \BibitemShut
  {NoStop}%
\bibitem [{\citenamefont {Hally}\ \emph {et~al.}(2012)\citenamefont {Hally},
  \citenamefont {Logan},\ and\ \citenamefont {Pilkington}}]{Hally:2012pu}%
  \BibitemOpen
  \bibfield  {author} {\bibinfo {author} {\bibfnamefont {K.}~\bibnamefont
  {Hally}}, \bibinfo {author} {\bibfnamefont {H.~E.}\ \bibnamefont {Logan}},\
  and\ \bibinfo {author} {\bibfnamefont {T.}~\bibnamefont {Pilkington}},\
  }\href {https://doi.org/10.1103/PhysRevD.85.095017} {\bibfield  {journal}
  {\bibinfo  {journal} {Phys. Rev. D}\ }\textbf {\bibinfo {volume} {85}},\
  \bibinfo {pages} {095017} (\bibinfo {year} {2012})},\ \Eprint
  {https://arxiv.org/abs/1202.5073} {arXiv:1202.5073 [hep-ph]} \BibitemShut
  {NoStop}%
\bibitem [{\citenamefont {Weinberg}(1979)}]{Weinberg:1979sa}%
  \BibitemOpen
  \bibfield  {author} {\bibinfo {author} {\bibfnamefont {S.}~\bibnamefont
  {Weinberg}},\ }\href {https://doi.org/10.1103/PhysRevLett.43.1566} {\bibfield
   {journal} {\bibinfo  {journal} {Phys. Rev. Lett.}\ }\textbf {\bibinfo
  {volume} {43}},\ \bibinfo {pages} {1566} (\bibinfo {year}
  {1979})}\BibitemShut {NoStop}%
\bibitem [{\citenamefont {Buchmuller}\ and\ \citenamefont
  {Wyler}(1986)}]{Buchmuller:1985jz}%
  \BibitemOpen
  \bibfield  {author} {\bibinfo {author} {\bibfnamefont {W.}~\bibnamefont
  {Buchmuller}}\ and\ \bibinfo {author} {\bibfnamefont {D.}~\bibnamefont
  {Wyler}},\ }\href {https://doi.org/10.1016/0550-3213(86)90262-2} {\bibfield
  {journal} {\bibinfo  {journal} {Nucl. Phys. B}\ }\textbf {\bibinfo {volume}
  {268}},\ \bibinfo {pages} {621} (\bibinfo {year} {1986})}\BibitemShut
  {NoStop}%
\bibitem [{\citenamefont {Leung}\ \emph {et~al.}(1986)\citenamefont {Leung},
  \citenamefont {Love},\ and\ \citenamefont {Rao}}]{Leung:1984ni}%
  \BibitemOpen
  \bibfield  {author} {\bibinfo {author} {\bibfnamefont {C.~N.}\ \bibnamefont
  {Leung}}, \bibinfo {author} {\bibfnamefont {S.}~\bibnamefont {Love}},\ and\
  \bibinfo {author} {\bibfnamefont {S.}~\bibnamefont {Rao}},\ }\href
  {https://doi.org/10.1007/BF01588041} {\bibfield  {journal} {\bibinfo
  {journal} {Z. Phys. C}\ }\textbf {\bibinfo {volume} {31}},\ \bibinfo {pages}
  {433} (\bibinfo {year} {1986})}\BibitemShut {NoStop}%
\bibitem [{\citenamefont {Grzadkowski}\ \emph {et~al.}(2010)\citenamefont
  {Grzadkowski}, \citenamefont {Iskrzynski}, \citenamefont {Misiak},\ and\
  \citenamefont {Rosiek}}]{Grzadkowski:2010es}%
  \BibitemOpen
  \bibfield  {author} {\bibinfo {author} {\bibfnamefont {B.}~\bibnamefont
  {Grzadkowski}}, \bibinfo {author} {\bibfnamefont {M.}~\bibnamefont
  {Iskrzynski}}, \bibinfo {author} {\bibfnamefont {M.}~\bibnamefont {Misiak}},\
  and\ \bibinfo {author} {\bibfnamefont {J.}~\bibnamefont {Rosiek}},\ }\href
  {https://doi.org/10.1007/JHEP10(2010)085} {\bibfield  {journal} {\bibinfo
  {journal} {JHEP}\ }\textbf {\bibinfo {volume} {10}},\ \bibinfo {pages}
  {085}},\ \Eprint {https://arxiv.org/abs/1008.4884} {arXiv:1008.4884 [hep-ph]}
  \BibitemShut {NoStop}%
\bibitem [{\citenamefont {Branco}\ \emph {et~al.}(2012)\citenamefont {Branco},
  \citenamefont {Ferreira}, \citenamefont {Lavoura}, \citenamefont {Rebelo},
  \citenamefont {Sher},\ and\ \citenamefont {Silva}}]{Branco:2011iw}%
  \BibitemOpen
  \bibfield  {author} {\bibinfo {author} {\bibfnamefont {G.}~\bibnamefont
  {Branco}}, \bibinfo {author} {\bibfnamefont {P.}~\bibnamefont {Ferreira}},
  \bibinfo {author} {\bibfnamefont {L.}~\bibnamefont {Lavoura}}, \bibinfo
  {author} {\bibfnamefont {M.}~\bibnamefont {Rebelo}}, \bibinfo {author}
  {\bibfnamefont {M.}~\bibnamefont {Sher}},\ and\ \bibinfo {author}
  {\bibfnamefont {J.~P.}\ \bibnamefont {Silva}},\ }\href
  {https://doi.org/10.1016/j.physrep.2012.02.002} {\bibfield  {journal}
  {\bibinfo  {journal} {Phys. Rept.}\ }\textbf {\bibinfo {volume} {516}},\
  \bibinfo {pages} {1} (\bibinfo {year} {2012})},\ \Eprint
  {https://arxiv.org/abs/1106.0034} {arXiv:1106.0034 [hep-ph]} \BibitemShut
  {NoStop}%
\bibitem [{\citenamefont {Aad}\ \emph {et~al.}(2020{\natexlab{b}})\citenamefont
  {Aad} \emph {et~al.}}]{ATLAS:2020wny}%
  \BibitemOpen
  \bibfield  {author} {\bibinfo {author} {\bibfnamefont {G.}~\bibnamefont
  {Aad}} \emph {et~al.} (\bibinfo {collaboration} {ATLAS}),\ }\href@noop {} {\
  (\bibinfo {year} {2020}{\natexlab{b}})},\ \Eprint
  {https://arxiv.org/abs/2004.03969} {arXiv:2004.03969 [hep-ex]} \BibitemShut
  {NoStop}%
\bibitem [{\citenamefont {Schmidt}(2016)}]{Schmidt_2016}%
  \BibitemOpen
  \bibfield  {author} {\bibinfo {author} {\bibfnamefont {B.}~\bibnamefont
  {Schmidt}},\ }\href {https://doi.org/10.1088/1742-6596/706/2/022002}
  {\bibfield  {journal} {\bibinfo  {journal} {Journal of Physics: Conference
  Series}\ }\textbf {\bibinfo {volume} {706}},\ \bibinfo {pages} {022002}
  (\bibinfo {year} {2016})}\BibitemShut {NoStop}%
\bibitem [{\citenamefont {Dulat}\ \emph {et~al.}(2018)\citenamefont {Dulat},
  \citenamefont {Lazopoulos},\ and\ \citenamefont
  {Mistlberger}}]{Dulat:2018rbf}%
  \BibitemOpen
  \bibfield  {author} {\bibinfo {author} {\bibfnamefont {F.}~\bibnamefont
  {Dulat}}, \bibinfo {author} {\bibfnamefont {A.}~\bibnamefont {Lazopoulos}},\
  and\ \bibinfo {author} {\bibfnamefont {B.}~\bibnamefont {Mistlberger}},\
  }\href {https://doi.org/10.1016/j.cpc.2018.06.025} {\bibfield  {journal}
  {\bibinfo  {journal} {Comput. Phys. Commun.}\ }\textbf {\bibinfo {volume}
  {233}},\ \bibinfo {pages} {243} (\bibinfo {year} {2018})},\ \Eprint
  {https://arxiv.org/abs/1802.00827} {arXiv:1802.00827 [hep-ph]} \BibitemShut
  {NoStop}%
\bibitem [{QCD(2015)}]{QCDEW2015}%
  \BibitemOpen
  \href@noop {} {\emph {\bibinfo {title} {Proc.\ QCD, EW, and tools at 100
  TeV}}},\ Introduction\ (\bibinfo  {publisher} {ATLAS},\ \bibinfo {address}
  {Geneva, Switzerland},\ \bibinfo {year} {2015})\BibitemShut {NoStop}%
\bibitem [{\citenamefont {de~Florian}\ \emph {et~al.}(2016)\citenamefont
  {de~Florian} \emph {et~al.}}]{deFlorian:2016spz}%
  \BibitemOpen
  \bibfield  {author} {\bibinfo {author} {\bibfnamefont {D.}~\bibnamefont
  {de~Florian}} \emph {et~al.} (\bibinfo {collaboration} {LHC Higgs Cross
  Section Working Group})\ }\textbf {\bibinfo {volume} {2/2017}},\ \href
  {https://doi.org/10.23731/CYRM-2017-002} {10.23731/CYRM-2017-002} (\bibinfo
  {year} {2016}),\ \Eprint {https://arxiv.org/abs/1610.07922} {arXiv:1610.07922
  [hep-ph]} \BibitemShut {NoStop}%
\bibitem [{\citenamefont {Batell}\ \emph {et~al.}(2015)\citenamefont {Batell},
  \citenamefont {McCullough}, \citenamefont {Stolarski},\ and\ \citenamefont
  {Verhaaren}}]{Batell:2015koa}%
  \BibitemOpen
  \bibfield  {author} {\bibinfo {author} {\bibfnamefont {B.}~\bibnamefont
  {Batell}}, \bibinfo {author} {\bibfnamefont {M.}~\bibnamefont {McCullough}},
  \bibinfo {author} {\bibfnamefont {D.}~\bibnamefont {Stolarski}},\ and\
  \bibinfo {author} {\bibfnamefont {C.~B.}\ \bibnamefont {Verhaaren}},\ }\href
  {https://doi.org/10.1007/JHEP09(2015)216} {\bibfield  {journal} {\bibinfo
  {journal} {JHEP}\ }\textbf {\bibinfo {volume} {09}},\ \bibinfo {pages}
  {216}},\ \Eprint {https://arxiv.org/abs/1508.01208} {arXiv:1508.01208
  [hep-ph]} \BibitemShut {NoStop}%
%%CITATION = ARXIV:1508.01208;%%
\bibitem [{\citenamefont {Falk}\ \emph {et~al.}(1995)\citenamefont {Falk},
  \citenamefont {Olive},\ and\ \citenamefont {Srednicki}}]{Falk:1995fk}%
  \BibitemOpen
  \bibfield  {author} {\bibinfo {author} {\bibfnamefont {T.}~\bibnamefont
  {Falk}}, \bibinfo {author} {\bibfnamefont {K.~A.}\ \bibnamefont {Olive}},\
  and\ \bibinfo {author} {\bibfnamefont {M.}~\bibnamefont {Srednicki}},\ }\href
  {https://doi.org/10.1016/0370-2693(95)00617-T} {\bibfield  {journal}
  {\bibinfo  {journal} {Phys. Lett. B}\ }\textbf {\bibinfo {volume} {354}},\
  \bibinfo {pages} {99} (\bibinfo {year} {1995})},\ \Eprint
  {https://arxiv.org/abs/hep-ph/9502401} {arXiv:hep-ph/9502401} \BibitemShut
  {NoStop}%
\bibitem [{\citenamefont {Contino}\ \emph {et~al.}(2017)\citenamefont {Contino}
  \emph {et~al.}}]{Contino:2016spe}%
  \BibitemOpen
  \bibfield  {author} {\bibinfo {author} {\bibfnamefont {R.}~\bibnamefont
  {Contino}} \emph {et~al.},\ }\href
  {https://doi.org/10.23731/CYRM-2017-003.255} {\bibfield  {journal} {\bibinfo
  {journal} {CERN Yellow Rep.}\ ,\ \bibinfo {pages} {255}} (\bibinfo {year}
  {2017})},\ \Eprint {https://arxiv.org/abs/1606.09408} {arXiv:1606.09408
  [hep-ph]} \BibitemShut {NoStop}%
\bibitem [{\citenamefont {Djouadi}(2008)}]{Djouadi:2005gj}%
  \BibitemOpen
  \bibfield  {author} {\bibinfo {author} {\bibfnamefont {A.}~\bibnamefont
  {Djouadi}},\ }\href {https://doi.org/10.1016/j.physrep.2007.10.005}
  {\bibfield  {journal} {\bibinfo  {journal} {Phys. Rept.}\ }\textbf {\bibinfo
  {volume} {459}},\ \bibinfo {pages} {1} (\bibinfo {year} {2008})},\ \Eprint
  {https://arxiv.org/abs/hep-ph/0503173} {arXiv:hep-ph/0503173 [hep-ph]}
  \BibitemShut {NoStop}%
%%CITATION = HEP-PH/0503173;%%
\bibitem [{\citenamefont {Carena}\ \emph {et~al.}(2012)\citenamefont {Carena},
  \citenamefont {Low},\ and\ \citenamefont {Wagner}}]{Carena:2012xa}%
  \BibitemOpen
  \bibfield  {author} {\bibinfo {author} {\bibfnamefont {M.}~\bibnamefont
  {Carena}}, \bibinfo {author} {\bibfnamefont {I.}~\bibnamefont {Low}},\ and\
  \bibinfo {author} {\bibfnamefont {C.~E.~M.}\ \bibnamefont {Wagner}},\ }\href
  {https://doi.org/10.1007/JHEP08(2012)060} {\bibfield  {journal} {\bibinfo
  {journal} {JHEP}\ }\textbf {\bibinfo {volume} {08}},\ \bibinfo {pages}
  {060}},\ \Eprint {https://arxiv.org/abs/1206.1082} {arXiv:1206.1082 [hep-ph]}
  \BibitemShut {NoStop}%
%%CITATION = ARXIV:1206.1082;%%
\end{thebibliography}%

\end{document}